\documentclass{aa}

\usepackage{graphicx}
\usepackage{natbib}
\usepackage{scalerel}
\usepackage{float}
\usepackage[table]{xcolor}
\usepackage{soul}
\usepackage{rotating}
\usepackage{txfonts}
\usepackage[pdfencoding=auto,psdextra]{hyperref}
\hypersetup{
    colorlinks=true,
    linkcolor=blue,
    filecolor=magenta,      
    urlcolor=blue,
    citecolor=blue
}
\makeatletter
\renewcommand*\aa@pageof{, page \thepage{} of \pageref*{LastPage}}
\makeatother

\usepackage[utf8]{inputenc}

\usepackage[switch, modulo]{lineno}

\usepackage{euclid}
\usepackage{orcidlink}
\usepackage{amsmath} 
\usepackage{amssymb}
\usepackage{mathtools}
\usepackage{multirow}
\usepackage{array}
\usepackage{makecell}
\usepackage{siunitx}
\usepackage{color}
\usepackage{graphicx}
\usepackage[caption=false]{subfig}
\usepackage{tabularx}
\usepackage{booktabs}
\usepackage{array}    
\usepackage{booktabs} 

\begin{document}
%
%

\title{Deep Learning galaxy cluster's  structural parameters from Weak Lensing observations}

   
\newcommand{\orcid}[1]{} 
\author{\normalsize
  M.~Fogliardi$^{1}$\orcidlink{0009-0006-4964-5311}\thanks{\email{fglmhl@unife.it}}, 
  M.~Meneghetti$^{2,3}$\orcidlink{0000-0003-1225-7084}, 
  C.~Giocoli$^{2,3}$\orcidlink{0000-0002-9590-7961},
  L.~Moscardini$^{4,2,3}$,
  P.~Rosati$^{1,2}$,
  L.~Leuzzi$^{2}$\orcidlink{0009-0006-4479-7017}, 
  G.~Angora$^{1,5}$,
  L.~Bazzanini$^{1,2}$\orcidlink{0000-0003-0727-013}, and
  C.~Spinelli$^{6}$}

\institute{$^{1}$ Department of Physics and Earth Science, University of Ferrara, via Saragat 1, I--44122, Ferrara, Italy\label{unife}\\
$^{2}$ INAF -- OAS, Osservatorio di Astrofisica e Scienza dello Spazio di Bologna, via Gobetti 93/3, I-40129 Bologna, Italy\label{inafbo} \\
$^{3}$ INFN-Sezione di Bologna, Viale Berti Pichat 6/2, 40127 Bologna, Italy \\
$^{4}$ Dipartimento di Fisica e Astronomia "Augusto Righi" - Alma
Mater Studiorum Università di Bologna, via Gobetti 93/2, 40129 Bologna, Italy \\
$^{5}$ INAF-Osservatorio Astronomico di Capodimonte, Via Moiariello 16, 80131 Napoli, Italy \\
$^{6}$ Argelander-Institut für Astronomie (AIfA), Universität Bonn, Auf dem Hügel 71, 53121 Bonn, Germany
}
\date{Received xxx; accepted xxx}

%
%
\abstract{Galaxy clusters are the most massive gravitationally bound structures in the Universe and serve as crucial probes of cosmic evolution and structure formation. The unprecedented data volume expected from upcoming wide-field imaging surveys demands efficient automated analysis techniques capable of processing observations of tens of thousands of galaxy clusters.
In this work, we present a comprehensive study exploiting Convolutional Neural Networks (CNNs) to infer cluster structural parameters from weak gravitational lensing observations. 
We implemented and optimized three CNN architectures (VGG-Net, Inception-v4, and Inception-ResNet-v2) using PyTorch to analyze reduced shear maps. Training was conducted on $75~000$ synthetic observations generated with the \texttt{MOKA} software, simulating semi-analytical mass distributions of galaxy clusters at $z = 0.25$. The networks were trained to simultaneously predict five key cluster parameters: virial mass ($M_{\rm vir}$), NFW concentration parameters ($c_{\rm NFW}$, $c_{\rm smooth}$), substructure count ($n_{\rm sub}$), and smooth component mass fraction ($f_{\rm smooth}$).
Performance evaluation on 5000 test clusters revealed excellent predictive accuracy for primary cluster properties. When realistic observational noise was introduced ($n_{\rm gal} = 30$ galaxies/arcmin$^2$, $\sigma_\epsilon = 0.3$), mass prediction accuracy remained robust (RMS $\sim 1.02 \times 10^{14}$ M$_\odot$/$h$, corresponding to a $\sim20$\% deviation). Concentration parameters proved to be remarkably stable against noise corruption, with VGG-22 maintaining the lowest RMS values ($c_{\rm NFW}=0.72$, $c_{\rm smooth}= 0.73$). However, substructure characterization presented significant challenges, with all models exhibiting systematic underprediction and elevated RMS values for $n_{\rm sub}$.
A comparative analysis with traditional reduced-tangential shear profile fitting demonstrated that the CNN approach achieved superior performance. For mass estimation, VGG-22 showed a negligible median underestimation of $-0.01$ dex ($\sim-2.3$\% in linear units) with $\log_{10}(M_{\rm vir})[{\rm M}_\odot/h] = 14.60$, while traditional fitting matched ground truth median of $14.61$. More significantly, concentration parameter estimation revealed substantial improvements: VGG-22 achieved $c_{\rm vir} = 4.60$ ($+2.9$\% overestimation) compared to traditional fitting's systematic underestimation ($c_{\rm vir} = 3.84$, $\sim14$\% underestimation).
Our results demonstrate that CNNs offer a viable and efficient alternative to conventional analysis methods, particularly beneficial for the automated processing of large survey datasets.
}
%
%
    \keywords{Gravitational Lensing: weak, Deep Learning, Galaxy Clusters}
%
%
   \titlerunning{ }
   \authorrunning{ }
   
   \maketitle
%
%
%
%
   
\section{\label{sc:Intro}Introduction}
Gravitational lensing is a powerful probe of the matter distribution in galaxy clusters. Observations of this phenomenon can be used to study the mass distribution of cosmic structures dominated by dark matter (DM) and to test models of cosmic structure formation \citep{1992ARA&A..30..311B}.

In particular, weak gravitational lensing induces both a distortion in the shapes of background galaxies (commonly referred to as \emph{shear}) and a magnification of their images, owing to the gravitational field of intervening massive objects and large-scale structure \citep{bartelmann2001weak}. In galaxy clusters, weak lensing typically produces coherent distortions of background-galaxy shapes at the level of up to a few tens of percent. The weak-lensing signal, extracted from these small but coherent distortions, therefore provides a direct measure of the projected mass distribution of galaxy clusters \citep{1993ApJ...404..441K}. In addition, lensing magnification affects the observed surface number density of background galaxies behind clusters and enhances their apparent fluxes, effectively increasing the accessible volume probed along the line of sight \citep{Broadhurst_1995}.

In classical parametric analyses, the weak-lensing mass is recovered by fitting the tangential shear profile. This approach provides a robust estimate of the projected matter distribution, but it is model-dependent \citep{okabe10b,okabe13,giocoli14,umetsu14,giocoli24,giocoli25}. Standard model fitting, typically based on Navarro-Frenk-White (NFW) \citep{Navarro_1996} or Einasto \citep{einasto65,navarro04} profiles, must carefully account for the degeneracy between mass and concentration, as well as for halo miscentering and the contribution of the two-halo term describing the transition between the cluster and the surrounding large-scale structure \citep{viola10,giocoli24b}. Although these parametric approaches are well established, they often require manual masking of the innermost regions to avoid contamination from the Brightest Cluster Galaxy (BCG), together with complex likelihood formulations to account for the non-Gaussian nature of shape noise.

The mass models inferred from weak and strong lensing have a wide range of applications \citep{Kneib_2011,Meneghetti_2013,Treu_2016,Umetsu_2020}, all aimed at improving our understanding of the physical processes that govern the growth and evolution of cosmic structures.

Upcoming wide-field imaging surveys, both from space and from the ground, such as those carried out by the \emph{Euclid} mission \citep{laureijs2011euclid, EuclidSkyOverview}, the \emph{Nancy Grace Roman} Space Telescope \citep{green2012}, and the \emph{Vera Rubin} Observatory Legacy Survey of Space and Time \citep[VRO/LSST;][]{LSST2019}, are expected to increase the number of observed clusters to several hundreds of thousands \citep{Sartoris_2016,2025bhargava}, thereby opening the way to statistical studies of large samples of lensing clusters \citep{Q1-SP057}. In this context, deep-learning (DL) techniques offer an efficient alternative to more classical and computationally expensive approaches, by automating feature extraction and enabling the rapid analysis of large imaging datasets. Unlike traditional one-dimensional tangential-shear profile fitting, which assumes azimuthal symmetry and compresses the two-dimensional shear field, convolutional neural networks (CNNs) can exploit the full information content of reduced-shear maps, including pixel-to-pixel correlations and the signatures of non-spherical mass distributions such as substructures.

Indeed, the application of DL to gravitational lensing is a rapidly expanding field, with significant progress in tasks ranging from the identification of strong-lens systems \citep{petrillo2017, angora2023, leuzzi2024, Q1-SP048} to the inference of cosmological parameters from weak-lensing data \citep{Ribli_2019, Fluri_2019}. In particular, CNNs can autonomously learn the most informative features for image classification and regression directly from the training data.

In this paper, we test the ability of different CNN architectures to infer the large-scale structural properties of galaxy clusters. The parameters considered here are the total virial mass ($M_{\rm vir}$), the concentration of the overall halo ($c_{\rm NFW}$) and of its smooth component ($c_{\rm smooth}$), the number of substructures within the halo ($n_{\rm sub}$), and the fraction of mass bound to the smooth halo component compared to the total mass ($f_{\rm smooth}$). This approach follows other successful applications of CNNs to the estimation of galaxy-cluster properties from lensing and other observational data \citep{Yan_2020, boruah2025diffusionbasedmassmapreconstruction,tominaga2026}.

We train our models on reduced-shear maps generated with the \texttt{MOKA} software \citep{giocoli12a}, which produces realistic semi-analytic mass distributions of galaxy clusters and computes the corresponding lensing quantities. In particular, we explore several architectures derived from VGGNet \citep{simonyan2015deep}, Inception-v4 \citep{szegedy2014going,szegedy2015rethinking,szegedy2016inceptionv4}, and Inception-ResNet-v2 \citep{he2015deep,szegedy2016inceptionv4}. These architectures have been widely used for image classification and have become standard benchmarks in the deep learning literature.

For each architecture, we implement several model variants to investigate the most suitable regularization and optimization strategies. Each model is trained on 75\,000 labelled shear maps to learn to predict the true values of the cluster's large-scale parameters. We also retrain our best-performing model on more realistic reduced-shear maps that include shape noise computed for a given lensed-galaxy number density. Finally, we compare the performance of the different models and discuss their respective properties.

This paper is organized as follows. In Sect.~\ref{sc:dset}, we describe the generation of the training set through weak-lensing simulations. In Sect.~\ref{sc:net}, we introduce the CNN frameworks and describe the architectures adopted in this work. In Sect.~\ref{sec:trainingthenetworks}, we present the training setup. In Sect.~\ref{sc:res}, we report the results obtained with the trained networks on both noiseless and noisy maps, and compare the predictions of our best-performing CNN with those derived from the traditional tangential-shear profile fitting method. Finally, in Sect.~\ref{sc:concl}, we summarize our findings and present our conclusions.

\section{\label{sc:dset} Dataset }

\subsection{The \texttt{MOKA} software}

In this paper, we analyze a dataset of weak lensing maps generated with the \texttt{MOKA} software \citep{giocoli12a}. This algorithm, starting from analytical prescriptions derived from numerical simulations and observations, creates surface mass density distributions for triaxial and substructured halos \citep{giocoli16b}. This approach is a valid alternative to full numerical hydrodynamical simulations for several applications, including gravitational lensing studies. Indeed, with a limited computational cost, it achieves a very high spatial resolution, which is necessary to resolve the inner structure of galaxies and clusters. In the simulations performed in this work, we assume isolated haloes, discarding the effects of correlated large-scale structures and any complex twisting or perturbations of the matter density distribution, unlike \citet{meneghetti16}. 

\texttt{MOKA}-generated galaxy clusters consist of large-scale components describing the cluster's smooth dark matter halos and a small-scale component accounting for dark matter subhalos and galaxies. We begin with the main ``smooth'' halos, whose  virial mass is defined as
\begin{equation}
M_{\rm vir}=\frac{4\pi}{3}\,r_{\rm vir}^3\,\Delta_{\rm vir}(z)\,\rho_{\rm c}(z),
\label{halomass}
\end{equation}
where $\rho_{\rm c}(z)$ is the critical density of the Universe at redshift $z$, and $\Delta_{\rm vir}(z)$ is the virial overdensity. Its matter distribution  is modeled with an NFW density profile \citep{Navarro_1996}, given by
\begin{equation}
    \rho(r) = \frac{\rho_s}{\left( \frac{r}{r_s}\right)\left(1+\frac{r}{r_s}\right)^2},
    \label{nfwprofile}
\end{equation}
where $r_s$ is the scale radius and $\rho_s$ is defined as the characteristic density of the DM halo.
The ratio between the virial and scale radii defines the halo concentration, $c_{\rm vir} \equiv r_{\rm vir}/r_s.$ 
Numerical simulations show that this parameter is correlated with halo mass, which defines an important relation in cosmology and is a prediction of the $\Lambda$CDM model. \texttt{MOKA} adopts the mass-concentration relation proposed by \cite{2009ApJ...707..354Z}, which links the concentration of a given halo with the time at which its main progenitor assembles 4\% of
its mass. Because of their tidal interaction with the surrounding density field during their collapse, DM halos are not spherical but triaxial. Such triaxiality is modeled in \texttt{MOKA} following \cite{2002ApJ...574..538J}. 

As the lensing signal is sensitive to the matter distribution in the central region of galaxy clusters ($r \sim 100$ kpc), we do assume a central galaxy adiabatically contracting the matter density distribution. Therefore, a BCG is added at the center of the halo using the halo occupation distribution (HOD) technique, which assumes that the stellar mass of a galaxy is tightly correlated with the depth of the potential well of the halo within which it formed \citep{Wang_2006}. For the BCG density profile a Hernquist \citep{1990ApJ...356..359H} profile is adopted:
\begin{equation}
    \rho_{\rm BCG} = \frac{\rho_g}{\left( \frac{r}{r_g}\right)\left(1+\frac{r}{r_g}\right)^3}.
\end{equation}
Here, $\rho_g$ and $r_g$ are the scale density taken from the Hernquist model and the scale radius related to the half-mass (or effective) radius, respectively.

The subhalo masses are drawn from the mass function proposed by \citet{Giocoli_2010}, based on the analysis of substructure abundances as a function of mass and redshift in the GIF2 cosmological $N$-body simulations \citep{2004MNRAS.355..819G}. This subhalo mass function is described by a power law with an exponential cutoff,
\begin{equation}
    \frac{ {\rm d} N_M}{{\rm d}\ln{m_{\rm sb}}} \equiv \frac{ m_{\rm sb}}{M_{\rm 0}} \frac{{\rm d}N}{{\rm d} m_{\rm sb}} = {N_{M_{\rm 0}}m_{\rm sb}}^\alpha \exp{(-\beta {\rm \xi^3})},
    \label{shmf}
\end{equation}
where $N_{M_{\rm 0}}$ denotes the normalization, $N_M$ the number of substructures in a given mass range, $m_{\rm sb}$ the subhalo mass, $M_{\rm 0}$ the host halo mass at $z=0$, and ${\xi}= m_{\rm sb}/M_{\rm 0}$. We adopt $\alpha=-0.9$ and $\beta \simeq 12.2715$. Following \cite{giocoli12a}, we assume that the subhalo radial distribution is described by 
\begin{equation}
    \frac{{n}(<x)}{ N} = \frac{{({ 1+a}c)}x^\beta}{({{ 1+a}c}x^\alpha)},
    \label{nsub}
\end{equation}
where $x$ is the distance to the host center in units of $r_{\rm vir}$, ${n}(<x)$ is the number of subhalos within $x$, $N$ is the total number of subhalos within $r_{\rm vir}$, ${a}=0.244$, $\alpha= 2$, $\beta=2.75$, and $c$ is the concentration of the host halo (related to $r_{\rm vir}$). The mass fraction contained in the subhalos with respect to the total mass of the halo is described by 
\begin{equation}
    f_{\rm sub} = \frac{\sum_i {{m}_{{\rm sb},i}}}{ M_0},
    \label{fsub}
\end{equation}
hence $f_{\rm smooth}=1-f_{\rm sub}$.
Since subhalos are assumed to host galaxies, we model thm as truncated Singular Isotermal Spheres (SISs), with density profiles given by \citep{2003ApJ...584..664K}:
\begin{equation}
\rho_{\rm sub}(r) = 
\begin{cases}
\begin{aligned}
& \frac{\sigma_v}{2 {\rm \pi G} r^2} & \text{for} & \quad r \leq {\rm R_{sub}} \\
& 0 & \text{for} & \quad  r > {\rm R_{sub}} \\
\end{aligned}
\end{cases},
\label{sis}
\end{equation}
where $\sigma_v$ is the velocity dispersion and ${\rm R_{sub}} = \frac{{{\rm G} m_{\rm sub}}}{2\sigma_v^2}$.

Finally, the total mass of the halos is given by the sum of smooth and clumpy components, and the BCG: ${M_{\rm vir}} = {M_{\rm smooth}}+ {\sum_{i=1}^{N_{\rm tot}} {m}_i} + {M_{\rm BCG}} $, where ${ m}_i$ is the mass of the $i-\rm th$ subhalo, $M_{\rm smooth}$ is the mass of the smooth halo and $M_{\rm BCG}$ is the mass of the BCG. 

Then, the lensing properties are derived from the 3D matter density of all components characterizing the halos. For each component, \texttt{MOKA} projects the density in a plane perpendicular to the line of sight and analytically obtains the projected mass density ${\rm \Sigma_{NFW}}$, ${\rm \Sigma_{sub}}$, ${\rm \Sigma_{BCG}}$. These quantities are then scaled with the appropriate ${\rm \Sigma_{cr}}$ value depending on the source and lens redshifts, assumed to be $z_{\rm s}=2$ and $z_{\rm l}=0.25$ respectively, and the total convergence map is obtained as 
\begin{equation}
     \kappa(x,y) = \kappa_{\rm DM}(x,y) + {\sum_{i=1}^{N_{\rm tot}} \kappa_{{\rm sub},i}(x-x_{\rm c}, y-y_{\rm c})} + \kappa_{\rm BCG}(x,y),
\end{equation}
where $x_c$ and $y_c$ denote the coordinates of the $i-\rm th$ substructure. Starting from the convergence map, the effective lensing potential $\phi$ can be obtained as
\begin{equation}
    \phi(x,y)= \frac{1}{\pi}\int_{\mathbb{R}^2}\kappa(\boldsymbol{\xi}')\ln|\boldsymbol{\xi}-\boldsymbol{\xi}'|\,\mathbf{d}^2\xi'.
\end{equation}
Finally, by calculating the derivatives of the effective lensing potential, the shear components are also obtained using
\begin{equation}
\gamma_1(x,y) = \frac{1}{2}\left(\frac{\partial^2 \phi}{\partial x^2} - \frac{\partial^2 \phi}{\partial y^2}\right)
\label{eq:shear1}
\end{equation}
and
\begin{equation}
\gamma_2(x,y) = \frac{\partial^2 \phi}{\partial x \partial y},
\label{eq:shear2}
\end{equation}
where $\gamma_1$ and $\gamma_2$ are the two components of the shear field. From the shear components, their reduced counterparts, which represent the real observable, can be obtained as
\begin{equation}
    g_i = \frac{\gamma_i}{1-\kappa},
\end{equation}
where $i=1,2$ represent the selected component and $g$ is the reduced shear. 
A comprehensive description of the convergence and shear map generation procedure is provided in Appendix~\ref{App:lensing_maps}. 

Figure~\ref{fig:maps} shows an example for the convergence and shear maps produced using the \texttt{MOKA} algorithm for a massive galaxy cluster ($ M_{\text{vir}} = 2.4 \times 10^{15}\ M_\odot/h$) acting as a lens at $z=0.25$.

\begin{figure*}
    \centering
    \includegraphics[width = 1\textwidth]{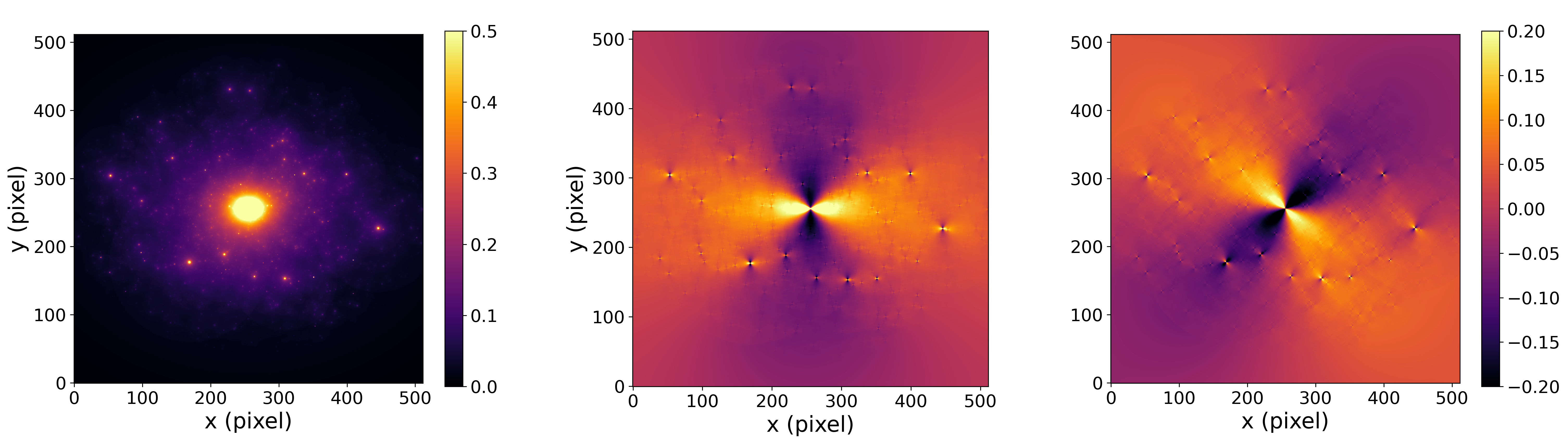}
    
    \caption{Example of simulated convergence (left panel) and shear component $\gamma_1$ and $\gamma_2$ (central and right panels, respectively)  maps obtained using the \texttt{MOKA} code for a massive cluster with total mass $ M_{\text{vir}} = 2.4 \times 10^{15}\ M_\odot/h$ at redshift $z=0.25$. The field of view (FOV) side length is $24.4$ arcminutes, and the pixel scale is $2.86$ arcsec.}
    \label{fig:maps}
\end{figure*}

\subsection{Noise simulation}
\label{sec:vggnoise}
Since noise affects lensing measurements, we also conducted a test using our best models to evaluate the CNN performances in more realistic conditions. To simulate realistic noise in actual observations, we adopt an observational configuration representative of current and upcoming Stage IV weak lensing surveys, such as the \textit{Euclid} mission, the \textit{Vera C. Rubin} Observatory (LSST), and the \textit{Nancy Grace Roman} Space Telescope. 
While these projects aim for effective source densities ranging from $\sim30$ to over $50$ galaxies per $arcmin^{-2}$, we adopt a fiducial baseline of $n_{\rm gal}= 30 \,(galaxies \times arcmin^{-2})$, consistent with the standard requirements for the \textit{Euclid} wide survey.

Then, we assume an intrinsic ellipticity distribution of the sources characterized by a dispersion of $\sigma_{\eta}= 0.3$. This intrinsic ellipticity dominates the distortion introduced by lensing, so we need to dilute the intrinsic term by averaging over a large number of sources, thereby making the lensing signal measurable. 

\begin{figure}
    \centering
    {\includegraphics[width=0.5\textwidth]{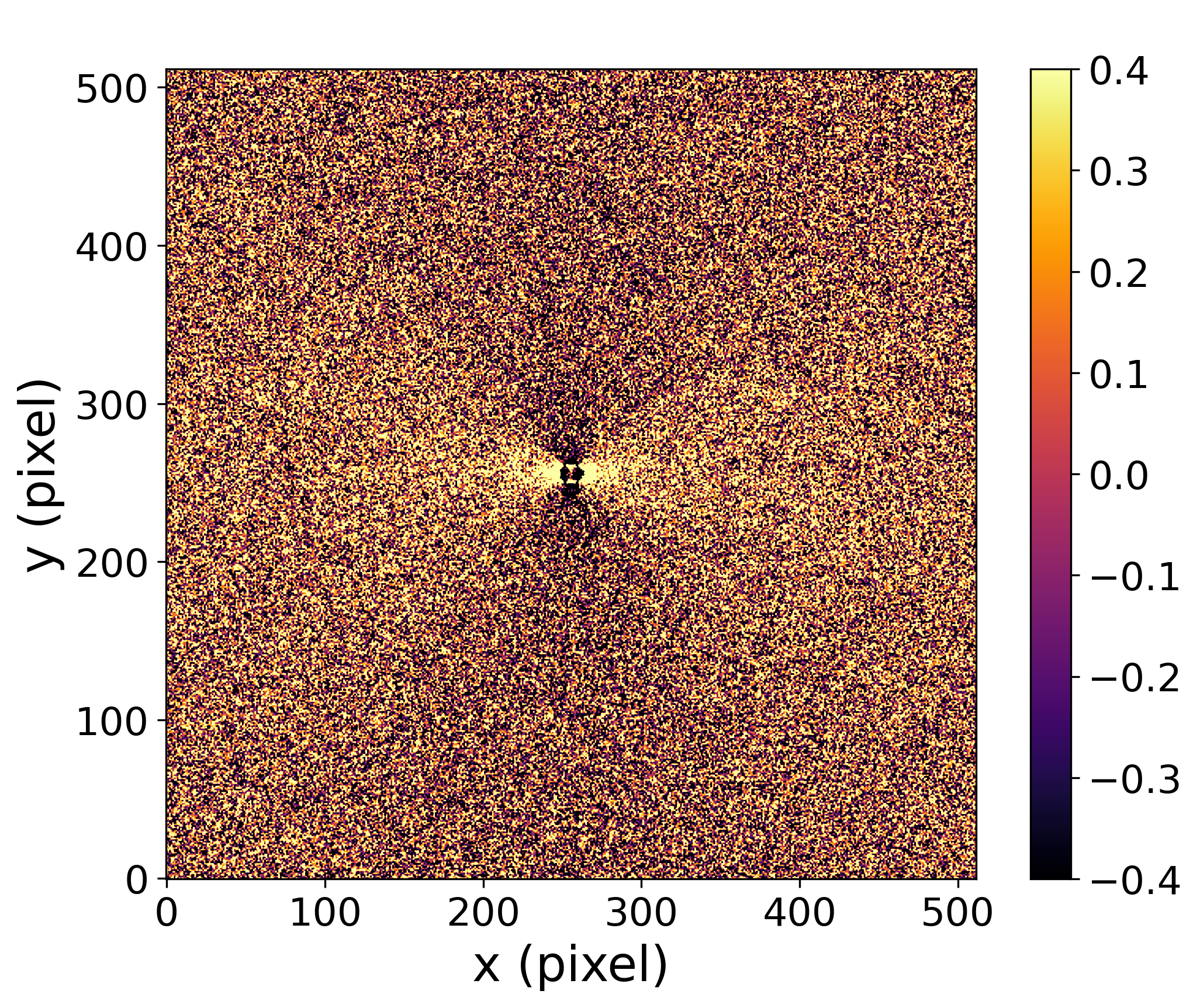}}\hfill
    {\includegraphics[width=0.5\textwidth]{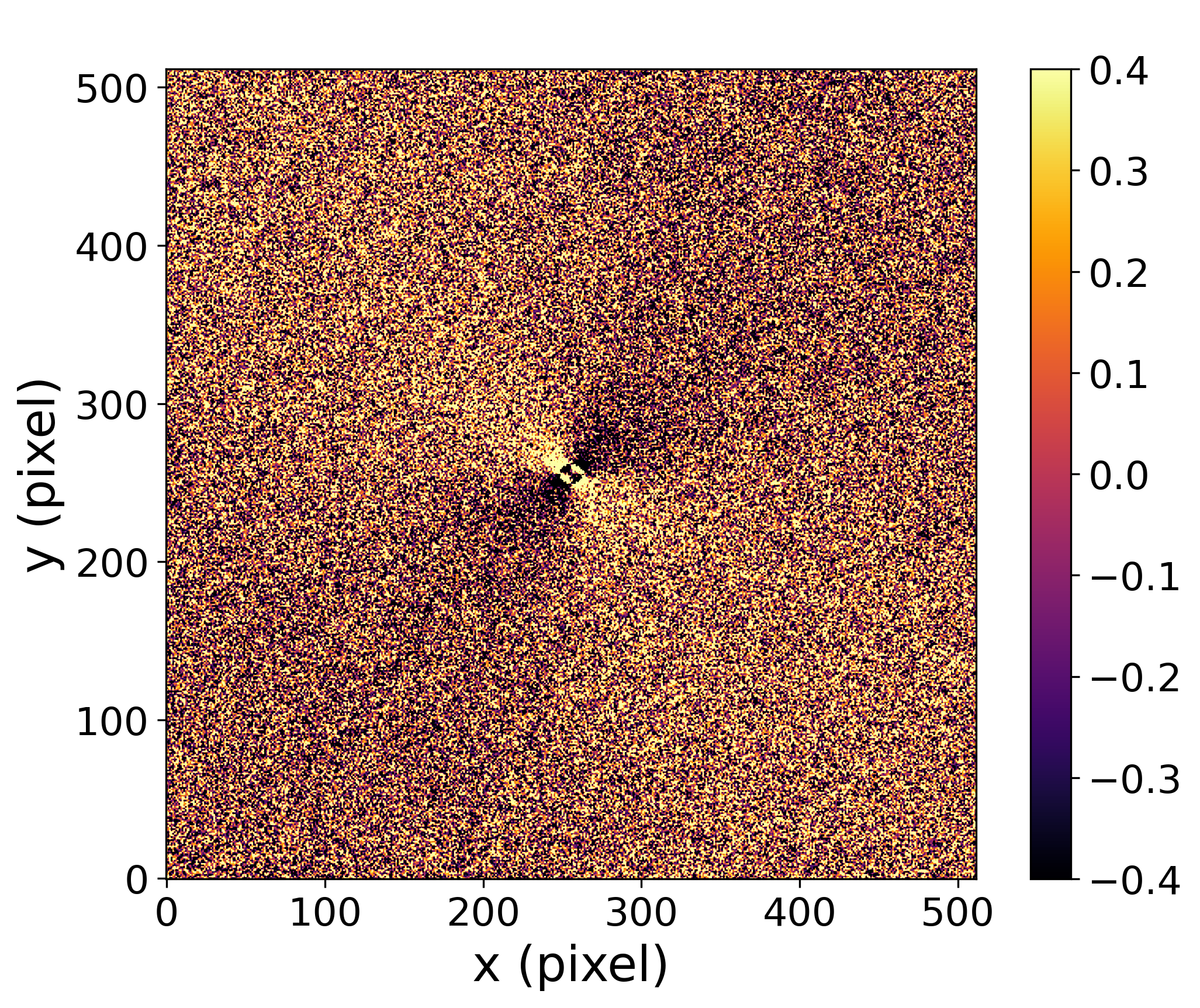}}\\
    \caption{Noisy reduced shear maps for a cluster at redshift $z=0.25$, showing the reduced shear component $g_1$ (upper panel) and $g_2$ (lower panel). This is the same cluster shown in Figure~\ref{fig:maps}.}
    \label{fig:noisemaps}
\end{figure}

Our shape noise per pixel is then calculated as described in \cite{Giocoli_2010}: 
\begin{equation}
    \sigma_{\rm noise, pixel} = \sqrt{\frac{\sigma_{\eta}^2}{(4\cdot {n_{\rm gal}}\cdot {p_{\rm size,arcmin}}^{2})}},
\end{equation}
where $p_{\rm size,arcmin}$ is the size of our pixel in arcminutes. We then generate noise on the reduced shear maps, assuming a Gaussian distribution, with zero mean and $\sigma = \sigma_{\rm noise, pixel}$. An example of the resulting noisy reduced shear maps is given in Figure~\ref{fig:noisemaps}.

\subsection{Dataset properties}

Weak lensing maps are stored as \texttt{FITS} (Flexible Image Transport System, \citealt{Pence_2010}) files, a digital file format useful for storing, transmitting, and processing data organized in multidimensional arrays (2D images, in this case).

Every cluster file has three extensions: one for the convergence and one for each of the two shear components. To feed our networks with appropriate training data, we first extract the labels from the headers, select only the parameters our networks need to predict, and store both the images and the labels in a single tensor. The selected labels are the halo virial mass $M_{\rm vir}$ (in units of $M_{\odot}/h$), the concentration of the NFW profile of the halo $c_{\rm NFW}$, the concentration of the NFW profile of the smooth component of the halo $ c_{\rm smooth}$, the number of subhalos $N_{\rm sub}$ and the smooth component mass fraction compared to the total halo mass, $f_{\rm smooth}$. 

The extraction process involves the entire dataset, which includes a catalog of 100\,000 simulated clusters at redshift $z = 0.25$. Then, the whole dataset is split into a training set of 75,000 maps, a validation set of 20,000 maps, and a test set for accuracy calculation, comprising the remaining 5,000 maps. 

\begin{figure*}
    \centering
    \includegraphics[width = 1\textwidth]{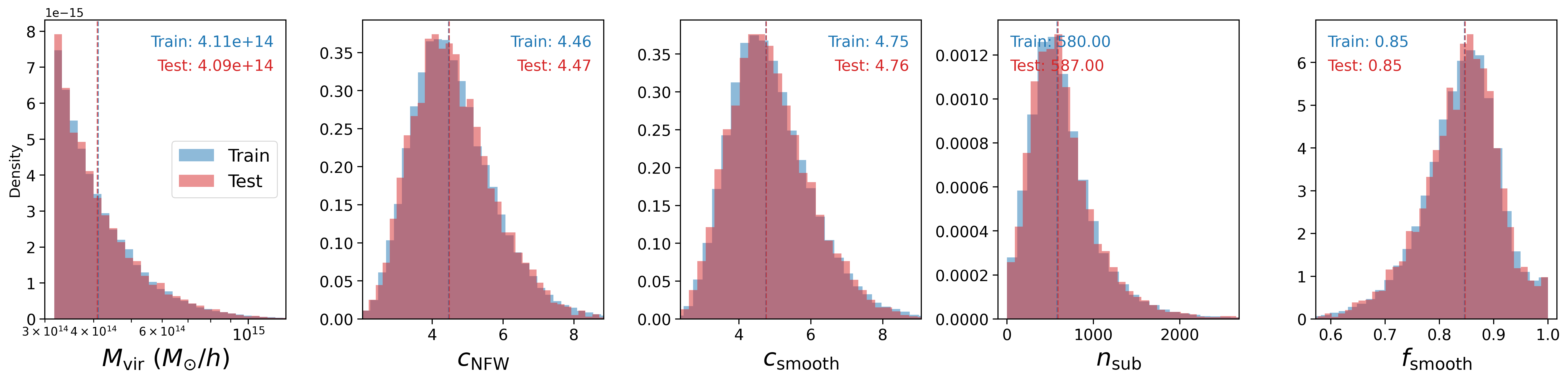}
    
    \caption{Distributions of cluster large-scale parameters in the training (blue) and test (red) datasets. Different panels from left to right correspond to the virial mass ($M_{\rm vir}$), the concentration of the entire halo ($c_{\rm NFW}$), the concentration of its smooth component ($c_{\rm smooth}$), the number of substructures within the halo ($n_{\rm sub}$), and the fraction of mass not contained in the substructures (i.e. the smooth component mass fraction, $f_{\rm smooth}$). Vertical dashed lines denote the median values for each distribution, also reported in each panel.}
    \label{fig:distr}
\end{figure*}

Figure~\ref{fig:distr} shows the distributions of the training and test set parameters, respectively. It is fairly straightforward to observe that these plots cover the same range of values, with very similar mean and median values (represented by red and blue dotted lines, respectively), indicating that the test set is a reliable depiction of the dataset. 

After label extraction and storage, the final step in data preprocessing is to normalize the distribution of each parameter by subtracting the mean and dividing by the standard deviation, ensuring values vary over a similar range. Normalizing the data helps bring the labels to a similar scale, preventing some features from dominating others during training. This uniformity aids the optimization algorithm's convergence, allowing the model to learn more efficiently and generalize better to unseen data by reducing the influence of biases in the input data \citep{lecun2012efficient}. Normalization also makes the optimization process more stable by preventing large gradients that could lead to oscillations or convergence issues, and reduces computational complexity by helping the model converge faster. This efficiency is particularly important when dealing with large datasets and complex architectures, as in this case \citep{goodfellow2016deep}. 

After normalizing the labels, we resize the images from their original resolution ($512\times512$ pixels) to match the input dimensions of our VGG-Net and Inception models ($224\times224$ and $299\times299$, respectively, for the initial experiment). However, for Inception models, we later reverted to the original $512\times512$ size after noticing an improved performance on full-sized images. Additionally, the images are normalized to have a zero mean and a standard deviation of \(\sigma = 1\), maintaining consistency with the preprocessing steps described above.

\section{\label{sc:net} Networks Architectures }

In this section, we introduce the network models used in this study. All of our models are developed using PyTorch \citep{NEURIPS2019_9015} and have been adapted from their original versions, as described below. Specifically, the models take the two reduced shear component maps derived from the \texttt{MOKA} simulations as input, with dimensions of $224\times224\times2$ for the VGG-inspired models and $512\times512\times2$ for the Inception ones. All models produce a $1\times5$ tensor as their output, consisting of five elements representing the predicted cluster parameters.

\subsection{VGG-Net}
\label{vgg2.2.1}

VGGNet is a deep CNN architecture designed for image classification, introduced by the Visual Geometry Group \citep{simonyan2015deep}. It employs a uniform architecture with stacked $3\times3$ convolutional filters (which is the smallest size to capture the notion of left/right, up/down, center), replacing larger receptive fields (e.g., $11\times11$ in \citealt{NIPS2012_c399862d}, $7\times7$ in \citealt{zeiler2013visualizing, sermanet2014overfeat}) to reduce parameters while increasing network depth. Inputs are fixed-size $224\times224$ RGB images, processed through convolutional layers ($\rm stride=1$, $\rm padding=1$, see Appendix \ref{sc:app A}), preserving spatial resolution. Max-pooling layers (2×2 window, $\rm stride=2$) intersperse the convolutional stacks, halving the input dimension. The final stage of the architecture employs Fully Connected (FC) layers with dropouts \citep{JMLR:v15:srivastava14a}, specifically adapted to optimize performance for our regression task. 

In the following, we describe the basic structure and underlying concepts of the models that we implemented in this work. 

\paragraph{VGG-19} This model adopts the configuration of the original VGG-16 architecture (16 layers; see Table~\ref{tab:vgg_variants}), with its substantial difference being the expansion of channel dimensions across all convolutional layers, which directly affects the total number of learnable parameters. To maintain the parameter reduction ratio in FC layers, three additional FC layers were incorporated, resulting in a 19-layer architecture.
\paragraph{VGG-22} This deeper model was later implemented to process native $512\times512$ resolution images. To mitigate parameter inflation while accommodating the increased input dimensions, three additional convolutional layers and a max-pooling layer were added before the FC stack (see Table~\ref{tab:vgg_variants}), extending the baseline VGG-19 architecture. However, empirical tests revealed superior performance on lower-resolution inputs ($224\times224$ pixels), prompting further optimization at this reduced scale. The final 22-layer model, while retaining a substantial number of trainable parameters, demonstrates improved predictive accuracy compared to alternative architectures.


\subsection{\label{sc:incnetowrk} Inception-v4}
Inception-v4 \citep{szegedy2016inceptionv4} is a computationally efficient CNN architecture designed for high-accuracy image recognition through hierarchical multiscale feature extraction. Its core innovation lies in optimized `Inception modules' (or `blocks', see Figure~\ref{fig:inc_blocks}a) that combine parallel convolutional pathways with varying receptive fields ($1\times1$, $3\times3$, and factorized $ \rm n\times1$/$ \rm 1\times n$ kernels), enabling simultaneous capture of fine-grained details and broader spatial patterns. 

The original architecture processes $299\times299$ RGB images through a preliminary feature extraction module named `stem', consisting of a series of strided convolutions and pooling operations (see Table~\ref{tab:stem_inc}). The stem is followed by a modular body comprising distinct sequential components:  
\begin{itemize}  
    \item Inception-A block: employs factorized $3\times3$ convolutions to balance feature diversity and parameter efficiency.  
    \item Inception-B block: utilizes asymmetric $1\times7$ and $7\times1$ convolutions for expanded receptive fields with minimal computational overhead.  
    \item Inception-C block: integrates hybrid filter combinations to optimize multiscale feature fusion.
    \item Dedicated A and B reduction modules between block groups, which combine max-pooling and parallel strided convolutions to halve feature map dimensions, avoiding representational bottlenecks.
\end{itemize}  
Regularization is enhanced by applying batch normalization \citep{ioffe2015batch} and label smoothing, which stabilize training dynamics and mitigate overfitting. The classifier substitutes traditional FC layers with global average pooling and dropout, followed by a softmax layer (omitted in our regression adaptation).

By decomposing large kernels into smaller sequential operations and optimizing grid reduction, Inception-v4 achieves state-of-the-art accuracy with 41\% fewer parameters than its predecessors \citep{szegedy2014going, ioffe2015batch, szegedy2015rethinking}, demonstrating superior efficiency in multiscale feature synthesis.

\subsection{\label{sc:resnet}Inception-Resnet-v2}
In the same work \citep{szegedy2016inceptionv4}, a second network based on the Inception architecture, called Inception-ResNet-V2, has been presented. This network differs from the previously presented Inception architecture in that it concerns the type of blocks involved in the convolutions and is based on residual learning. Residual Networks (ResNets), introduced by \cite{he2015deep}, address the degradation problem inherent to deep NNs, where increasing depth leads to higher training and test errors through residual learning. Unlike conventional architectures that directly learn target mappings $H(\mathbf{x})$, ResNets decompose the problem into learning residual functions $F(\mathbf{x}) = H(\mathbf{x}) - \mathbf{x}$. This is achieved via shortcut connections that bypass one or more convolutional layers, enabling identity mapping through the additive operation $\mathbf{y} = F(\mathbf{x}) + \mathbf{x}$. When input and output dimensions mismatch, a linear projection $W_s$ adjusts the shortcut path: $\mathbf{y} = F(\mathbf{x}) + W_s \mathbf{x}$. By simplifying the learning objective to incremental residuals, ResNets stabilize gradient flow and enable training of networks exceeding 1000 layers \citep{he2016identity}, which was unfeasible with traditional sequential architectures like VGGNet or Inception-v4 due to vanishing gradients \citep{279181, 2012arXiv1211.5063P}. The hybrid Inception-ResNet-v2 architecture merges this paradigm by integrating residual shortcuts into Inception modules (Figure~\ref{fig:inc_blocks}b). Here, the parallel convolutional pathways of Inception-v4 are augmented with additive skip connections, enabling simultaneous multiscale feature extraction and gradient stabilization. This synthesis retains Inception-v4’s stem and reduction modules to avoid representational bottlenecks during downsampling, but replaces plain Inception blocks with residual counterparts.

The architectural composition and theoretical basis of the developed models are elaborated below.

\paragraph{Inception models}
\label{par:incmodels}
This study systematically evaluated multiple configurations of Inception-v4 and Inception-ResNet-v2. Initial implementations faithfully replicated the original frameworks presented in \cite{szegedy2016inceptionv4}, processing 
$299\times299$ input images. Notably, upscaling the input dimensions to the original size ($512\times512$ pixels) yielded substantial accuracy improvements, contrary to the behavior observed in our VGG-based models. In order to accommodate larger inputs, the stem structure of Inception-v4 was modified (Table~\ref{tab:stem_inc}), while the ResNet variant retained its original configuration (Table~\ref{tab:stem_res}).
Final architectures (Table~\ref{tab:incres}) employ distinct block configurations: Inception-v4 integrates 3A/5B/3C blocks, whereas Inception-ResNet-v2 utilizes 5A/10B/5C blocks. FC layers were streamlined by progressively reducing channel counts from 1536 to 5. Dropout layers were incorporated after the first three FC layers, and softmax activation was omitted for regression compatibility. The optimized Inception-v4 and Inception-ResNet-v2 models contain 36.5 million and 35.8 million trainable parameters, respectively.

\section{Training the networks}
\label{sec:trainingthenetworks}
Our model training was performed on a NVIDIA Titan Xp GPU with 12 GB of memory and 3840 CUDA cores. Employing a GPU substantially reduces computation time and speeds up the training process of artificial NNs. 

Before training the models, it is essential to define a loss function, which measures the disparity between the network's
predictions and the correct output. In the context of this work, we selected the Mean Squared Error (MSE) loss function, defined as:
\begin{equation}
\text{MSE} = \frac{1}{n} \sum_{i=1}^{n} (\textbf{y}_i - \hat{\textbf{y}}_i)^2 \;,
\label{eqloss}
\end{equation}
where $n$ is the total number of training examples, $\textbf{y}_i$ and $\hat{\textbf{y}}_i$ are the predicted output value and the real output value, respectively. 

The purpose of the training process is to find the best combination of weights and biases (which are layer-specific updatable parameters) that minimizes the loss function. Usually, each iteration of the training process, known as an 'epoch', has an associated loss value, and its trend over time is checked to evaluate the effectiveness of the learning process, as the loss value is expected to decrease with the number of epochs.

Along with the loss function, we also need to choose an optimization method to train a network, which defines how the weights and biases are updated. Optimization is usually left to a set of pre-built optimizers, since differences in weight adjustment methods affect the network's accuracy. Our chosen optimizer is an improved version of the Adaptive Moment estimation (Adam, \citealt{kingma2017adam, reddi2019convergence}) called AdamW \citep{loshchilov2019decoupledweightdecayregularization}. In particular, Adam is an improved version of the Stochastic Gradient Descent algorithm (SGD, \citealt{goodfellow2016deep}). SGD updates the model weights $\theta_i$ by minimizing the loss function $\rm L(\theta)$ as described by the equation:
\begin{equation}
    \theta_{i+1} = \theta_i - \alpha \frac{\partial \rm L(\theta_i)}{\partial \theta_i},
\end{equation}
 where $\alpha$ represents the learning rate, a hyperparameter that defines the step length of each update in the negative gradient direction. The learning rate is manually tuned to avoid local minima and ensure convergence.
 
 Adam is an adaptive learning rate SGD optimization algorithm that combines the benefits of both Momentum Optimization \citep{sgdmomentum} and Root Mean Squared Propagation (RMSProp, \citealt{rmsprop}). Momentum Optimization helps the gradient descent algorithm converge faster and more reliably by adding a term that incorporates past weight updates, leading to more stable and efficient updates toward the optimal solution. On the other hand, RMSprop is an unpublished learning rate optimizer that adjusts the learning rate adaptively for each parameter based on the magnitudes of recent gradients. Additionally, Adam incorporates bias correction to counteract the effects of initial learning rate bias and momentum, resulting in more stable and reliable convergence. AdamW modifies Adam's implementation of weight decay \citep{goodfellow2016deep}, by decoupling it from the gradient update. \cite{adamw} shows that AdamW substantially improves Adam’s generalization performance.
 
 In all epochs, we train all weights across all layers using a learning rate scheduler that starts at $\alpha = 10^{-5}$ and decreases it by a factor of $0.5$ every 20 epochs. This approach facilitates deeper learning and more precise weight adjustment, while reducing the risk of overfitting \citep{goodfellow2016deep}. The AdamW optimizer was initialized with an exponential decay rate for the first- and second-moment estimates, with $\beta = (0.9, 0.999)$, and a weight decay coefficient of $10^{-2}$. Furthermore, to prevent overfitting, we include early stopping regularization \citep{prechelt1997, raskutti2011}.

\section{\label{sc:res} Results }
In this section, we present the training results and evaluate the performance of our neural networks on a test set of 5~000 clusters. The results are presented in the mass bins reported in Table~\ref{tab:bins}.
\renewcommand{\arraystretch}{1.2}
\begin{table}[H]
    \centering
    \caption{Bin distribution of clusters in the test set.}
    \begin{tabular}{ll}
        \hline\hline
        $ \log_{10}(M_{\text{vir}} \ [M_\odot/h]) $ & $N_{\rm obj}$ \\ 
        \hline
        $14.4\mbox{-}14.6$ & 2324 \\ 
        $14.6\mbox{-}14.8$ & 2013 \\ 
        $14.8\mbox{-}15.0$ & 565 \\
        $15.0\mbox{-}15.3$ & 98 \\
        \hline\hline
    \end{tabular}
    \tablefoot{Bins are defined for $ \log_{10}(M_{\text{vir}} \ [M_\odot/h]) $, $N_{\rm obj}$ refers to the total number of objects in each bin.}
    \label{tab:bins}
\end{table}
We begin by describing the traditional ML metrics used for regression tasks. Next, we assess models' performances on noiseless reduced shear maps, followed by an analysis of results obtained on reduced shear maps including simulated Euclid-like noise.

\subsection{Statistical estimators}
The traditional techniques for error analysis are built on the Bayesian framework, which provides a formal method for updating a model's state of knowledge and making predictions as new evidence becomes available. However, applying this to high-dimensional models like CNNs is computationally expensive, as Bayesian inference scales with the number of parameters. Furthermore, CNNs typically rely on gradient-based optimization (e.g., stochastic gradient descent) to learn from data, which may not align directly with the probabilistic framework of Bayesian inference. Therefore, we conduct a statistical error analysis of our networks using metrics such as bias and Root Mean Squared Error (RMSE). 

Let us define $y$ and $\hat{y}$ as the predicted output value array and the real output value array, respectively, with their difference given by $\Delta y = y-\hat{y}$. Assuming $N$ is the total number of test examples, we define:
\begin{itemize}
    \item the bias, as \begin{equation}
{\rm Bias}(\hat{y}) = E(\hat{y}) - y  = \frac{1}{N} \sum_{i=1}^{N}  \Delta {y_{i}} , 
\end{equation} 
which refers to the mean difference between the expected value of an estimator $\text{E}(\hat{y})$ and the true value of the parameter being estimated. In the context of CNNs, bias measures how accurately the model's predictions average across true values.
\item the standard deviation $\sigma$, as \begin{equation}
\sigma = \sqrt{\frac{1}{N} \sum_{i=1}^{N} (\Delta y_i - {\langle \Delta{y} \rangle})^2},
\end{equation} which measures the dispersion of a dataset relative to its mean. In the context of CNNs, $\sigma$ quantifies the spread of errors around the mean (the bias).
\item the Root Mean Squared error (RMS), as \begin{equation}
    \text{RMS} = \sqrt{\frac{1}{N} \sum_{i=1}^{N} (\Delta y_{i}^2)},
\end{equation} 
which measures the average magnitude of the errors. It is useful for evaluating a predictive model's overall performance.
\end{itemize}

\subsection{Results on noiseless maps}
The results obtained by analyzing \texttt{MOKA}-simulated noiseless shear maps are reported in Table~\ref{tab:cnn_performance_noiseless}. The best performances achieved by our CNN models were obtained using the VGG-19 architecture trained on images resized to $224 \times 224$ pixels, and the Inception-based models introduced in Sect.~\ref{par:incmodels}, trained on $512 \times 512$ pixel inputs. A comprehensive description of the network architectures is provided in Appendices~\ref{sc:app:vgg_arch} and~\ref{sc:app:inception}. 

\begin{figure}
    \centering
    {\includegraphics[width=0.5\textwidth]{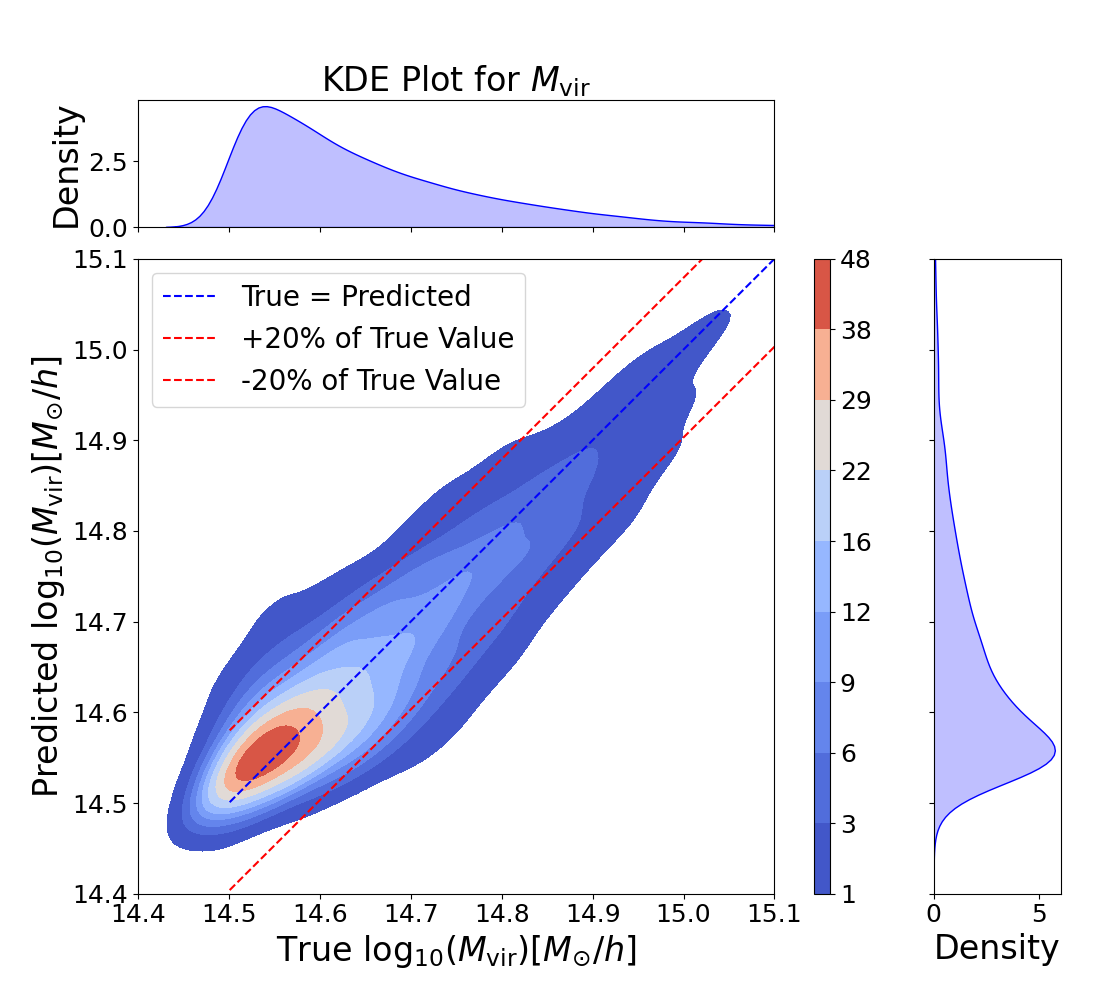}}\hfill
    {\includegraphics[width=0.5\textwidth]{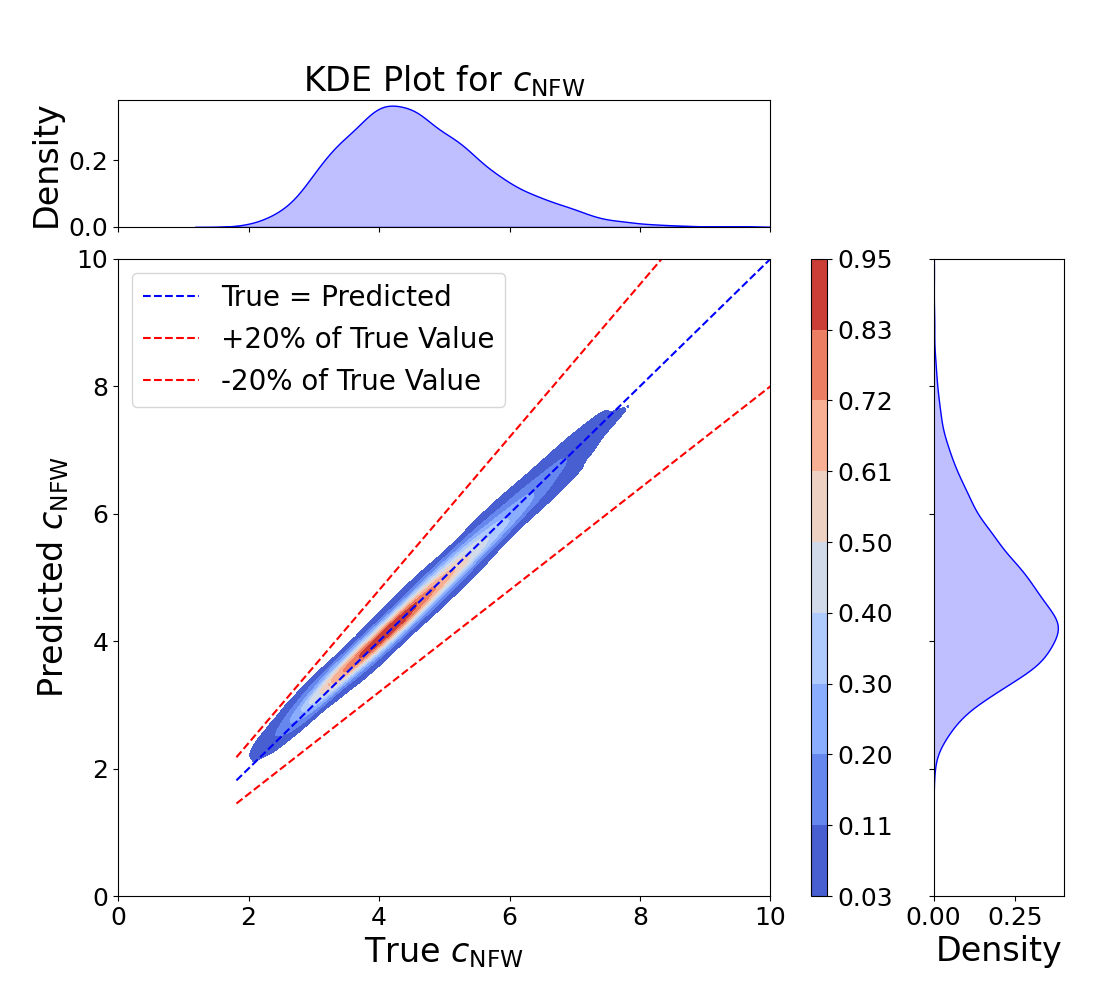}}\\
    \caption{KDE plots showing predicted vs target values for $M_{\rm vir}$ (upper panel) and $c_{\rm NFW}$ (lower panel) obtained applying our Resnet-V2 model on noiseless reduced shear maps of clusters located at $z = 0.25$.}
    \label{fig:kdeplotresnetnoiseless_m_c}
\end{figure}

The models exhibit a comparable overall precision in estimating the virial mass $ M_{\rm vir}$, with low biases and a similar overall RMS ($\sim0.7\times 10^{14}~ M_\odot/h$) across the entire mass range. However, performance degrades with increasing mass: for $\log_{10}(M_{\rm vir}) \in [15.0, 15.3]$, the absolute values of the bias increases substantially (up to $\sim1.3\times 10^{14}~M_\odot/h$, almost an order of magnitude larger than measured for $\log_{10}(M_{\rm vir}) \lesssim 14.8$), particularly for VGG-Net and Inception, indicating a systematic underestimation at the high-mass end. ResNet shows lower bias and RMS in this regime, suggesting a better capacity to generalize across mass scales (see Figure~\ref{fig:kdeplotresnetnoiseless_m_c}).

\begin{figure*}
    \centering
    \includegraphics[width = 1\textwidth]{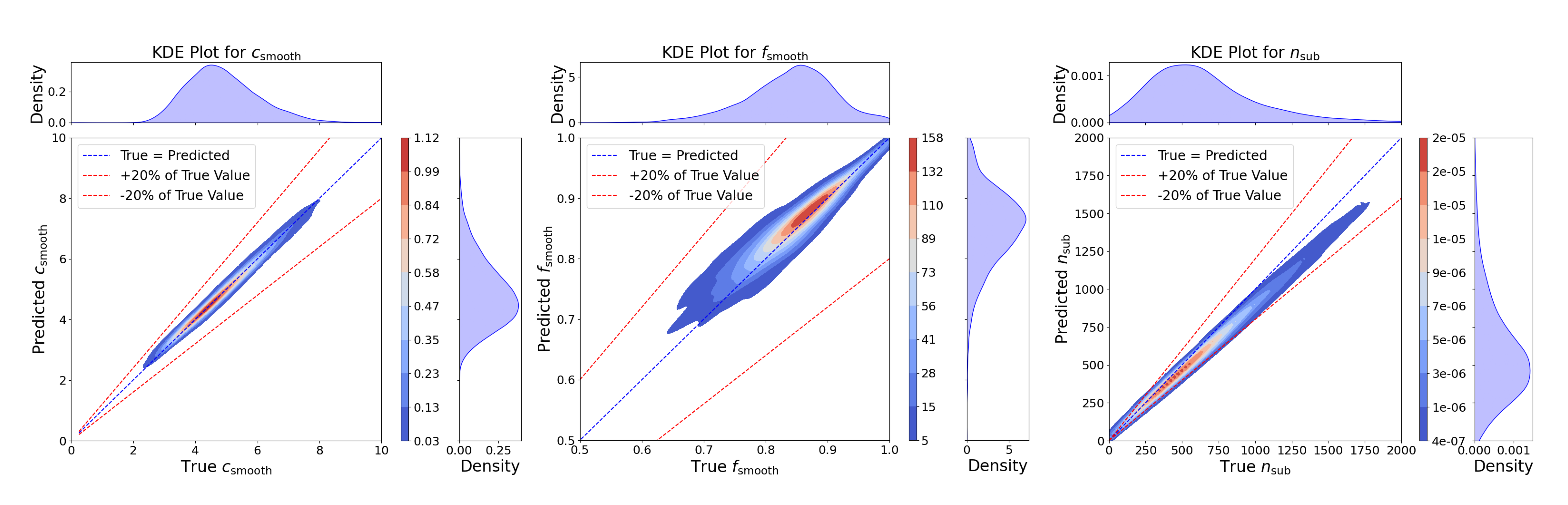}
    
    \caption{Same as Figure~\ref{fig:kdeplotresnetnoiseless_m_c} but for $c_{\rm smooth}$ (left panel), $f_{\rm smooth}$ (center panel)  and $n_{\rm sub}$ (right panel).}
    \label{fig:kdeplotsresnetnoiselesspanel}
\end{figure*}

For the concentration parameters $c_{\rm NFW}$ and $c_{\rm smooth}$, all models maintain low bias (<~$0.1$) and relatively small overall RMS values ($\sim0.2$), with ResNet again showing better performances. In particular, ResNet consistently yields the lowest biases and RMS for both concentration parameters, reflecting its robustness in capturing shape-related features and recovering halo profiles (see Figure~\ref{fig:kdeplotresnetnoiseless_m_c} and Figure~\ref{fig:kdeplotsresnetnoiselesspanel}).

The number of subhalos $n_{\rm sub}$ represents a more challenging regression target, with high bias and RMS values (notably 86 for ResNet and $\sim 60$ for other architectures, corresponding to a median $\sim 15\%~\text{-}~11\%$ error on the total respectively), showing a systematic underestimation bias for higher values of $n_{\rm sub}$ (see Figure~\ref{fig:kdeplotsresnetnoiselesspanel}) and indicating reduced sensitivity to small-scale features. This may reflect insufficient resolution in shear maps due to the resizing of the original image (from $512\times512$ to $224\times224$).

In contrast, the smooth component mass fraction ($f_{\rm smooth}=1-f_{\rm sub}$) is predicted with relatively low RMS ($0.024~\text{-}~0.030$) and biases (up to a maximum of $\sim2.5\%$ for Inception) across all models, showing that the global substructure content is more learnable than discrete subhalo counts.

Overall, ResNet offers slightly better performance on most parameters, suggesting that its deeper architecture may be better suited to capture complex shear patterns. VGG and Inception perform similarly across mass bins, implying that model complexity alone (e.g., Inception’s multiscale filters) does not guarantee gains for shear map regression. However, differences among the models are modest, and the predictions remain reliable across a range of cluster properties under noiseless conditions.

\renewcommand{\arraystretch}{1.2}
\begin{table*}[htbp]
\centering
\caption{Performance of CNN models on noiseless shear maps across mass bins and cluster parameters.}
\label{tab:cnn_performance_noiseless}
\begin{tabular}{lll|ccc}
\hline\hline
\textbf{Par.} & \textbf{Bin} & \textbf{Metric} & \textbf{VGG-19} & \textbf{Inception} & \textbf{ResNet} \\
\hline
\multirow{5}{*}{$\rm M_{\text{vir}}$} 
 & $14.4\mbox{-}14.6$ & B $\pm$ $\sigma$ & $0.25(7.1\%) \pm 0.33$ & $0.21(6.0\%) \pm 0.39$ & $0.19(5.4\%) \pm 0.39$ \\
 & $14.6\mbox{-}14.8$ & B $\pm$ $\sigma$ & $-0.19(-4.0\%) \pm 0.63$ & $-0.22(-4.7\%) \pm 0.65$ & $-0.16(-3.4\%) \pm 0.69$ \\
 & $14.8\mbox{-}15.0$ & B $\pm$ $\sigma$ & $-0.73(-10.0\%) \pm 0.97$ & $-0.70(-9.6\%) \pm 0.99$ & $-0.55(-7.5\%) \pm 1.05$ \\
 & $15.0\mbox{-}15.3$ & B $\pm$ $\sigma$ & $-1.27(-11.3\%) \pm 1.29$ & $-1.31(-11.7\%) \pm 1.37$ & $-0.96(-8.5\%) \pm 1.47$ \\
 & All & RMS & 0.69(14.1\%) & 0.71(14.5\%) & 0.70(14.3\%) \\
\hline
\multirow{5}{*}{$c_{\text{NFW}}$} 
 & $14.4\mbox{-}14.6$ & B $\pm$ $\sigma$ & $0.03(0.7\%) \pm 0.2$ & $-0.05(-1.1\%) \pm 0.18$ & $0.03(0.7\%) \pm 0.17$ \\
 & $14.6\mbox{-}14.8$ & B $\pm$ $\sigma$ & $0.01(0.2\%) \pm 0.2$ & $-0.07(-1.6\%) \pm 0.2$ & $0.00(0.0\%) \pm 0.19$ \\
 & $14.8\mbox{-}15.0$ & B $\pm$ $\sigma$ & $0.01(0.2\%) \pm 0.19$ & $-0.08(-1.9\%) \pm 0.18$ & $0.00(0.0\%) \pm 0.17$ \\
 & $15.0\mbox{-}15.3$ & B $\pm$ $\sigma$ & $0.04(1.0\%) \pm 0.24$ & $-0.05(-1.2\%) \pm 0.23$ & $0.02(0.5\%) \pm 0.22$ \\
 & All & RMS & 0.20(4.5\%) & 0.20(4.5\%) & 0.18(4.0\%) \\
\hline
\multirow{5}{*}{$c_{\text{smooth}}$} 
 & $14.4\mbox{-}14.6$ & B $\pm$ $\sigma$ & $0.01(0.2\%) \pm 0.19$ & $-0.12(-2.5\%) \pm 0.17$ & $0.01(0.2\%) \pm 0.16$ \\
 & $14.6\mbox{-}14.8$ & B $\pm$ $\sigma$ & $0.01(0.2\%) \pm 0.17$ & $-0.11(-2.3\%) \pm 0.18$ & $0.00(0.0\%) \pm 0.17$ \\
 & $14.8\mbox{-}15.0$ & B $\pm$ $\sigma$ & $0.02(0.4\%) \pm 0.17$ & $-0.12(-2.6\%) \pm 0.16$ & $-0.01(-0.2\%) \pm 0.15$ \\
 & $15.0\mbox{-}15.3$ & B $\pm$ $\sigma$ & $0.06(1.3\%) \pm 0.22$ & $-0.10(-2.2\%) \pm 0.2$ & $0.00(0.0\%) \pm 0.21$ \\
 & All & RMS & 0.18(3.8\%) & 0.21(4.4\%) & 0.16(3.4\%) \\
\hline
\multirow{5}{*}{$n_{\text{sub}}$} 
 & $14.4\mbox{-}14.6$ & B $\pm$ $\sigma$ & $-6(-1.3\%) \pm 43$ & $-24(-5.1\%) \pm 41$ & $-37(-7.8\%) \pm 43$ \\
 & $14.6\mbox{-}14.8$ & B $\pm$ $\sigma$ & $-23(-3.6\%) \pm 53$ & $-37(-5.7\%) \pm 51$ & $-60(-9.3\%) \pm 54$ \\
 & $14.8\mbox{-}15.0$ & B $\pm$ $\sigma$ & $-61(-6.1\%) \pm 77$ & $-67(-6.7\%) \pm 68$ & $-106(-10.7\%) \pm 79$ \\
 & $15.0\mbox{-}15.3$ & B $\pm$ $\sigma$ & $-154(-10.3\%) \pm 158$ & $-111(-7.4\%) \pm 142$ & $-202(-13.6\%) \pm 154$ \\
 & All & RMS & 65(11.1\%) & 66(11.3\%) & 86(14.7\%) \\
\hline
\multirow{5}{*}{$f_{\text{smooth}}$} 
 & $14.4\mbox{-}14.6$ & B $\pm$ $\sigma$ & $0.8(0.9\%) \pm 1.9$ & $2.5(2.9\%) \pm 1.9$ & $1.5(1.8\%) \pm 2.0$ \\
 & $14.6\mbox{-}14.8$ & B $\pm$ $\sigma$ & $-0.3(-0.4\%) \pm 2.5$ & $1.3(1.5\%) \pm 2.4$ & $0.6(0.7\%) \pm 2.5$ \\
 & $14.8\mbox{-}15.0$ & B $\pm$ $\sigma$ & $-0.6(-0.7\%) \pm 2.7$ & $0.9(1.1\%) \pm 2.7$ & $0.3(0.4\%) \pm 2.8$ \\
 & $15.0\mbox{-}15.3$ & B $\pm$ $\sigma$ & $-0.4(-0.5\%) \pm 2.9$ & $0.8(1.0\%) \pm 2.8$ & $0.5(0.6\%) \pm 2.9$ \\
 & All & RMS & 2.4(2.8\%) & 3.0(3.5\%) & 2.6(3.1\%) \\
\hline
\hline\hline
\end{tabular}
\tablefoot{Metrics include Bias (B), standard deviation ($\sigma$) and Root Mean Squared error (RMS). Bins are defined for $ \log_{10}(M_{\text{vir}} \ [M_\odot/h]) $; the RMS is evaluated over the full range. All $M_{\text{vir}}$ metrics are given in units of [$10^{14}~M_\odot/h$]; metrics for $f_{\text{smooth}}$ are given in \% units. Percentages in parentheses are computed relative to the median of the corresponding true parameter in the same mass bin for the bias, and relative to the full-sample median for the RMS row. For the $M_{\rm vir}$ RMS percentages, the dex scatter is obtained and then turned into a percent value.}
\end{table*}

\subsection{Results on noisy maps}

We now assess the performance of the networks on noisy reduced shear maps, whose statistical results are summarized in Table~\ref{tab:cnn_performance_noisy}. Adding shape noise degrades predictive accuracy across all architectures and parameter settings, as expected, but distinct trends emerge depending on the model and mass regime.

In this framework, our best results were obtained using the VGG-22 model on $224\times224$ images. For the Inception and ResNet models, the results on noisy maps presented in Table~\ref{tab:cnn_performance_noisy} were obtained using the same architecture configurations as in the noiseless case.

\renewcommand{\arraystretch}{1.2}
\begin{table*}[htbp]
\centering
\caption{Performance of CNN models on noisy (N) reduced shear maps across mass bins and cluster parameters.}
\label{tab:cnn_performance_noisy}
\begin{tabular}{lll|ccc}
\hline\hline
\textbf{Par.} & \textbf{Bin} & \textbf{Metric} & \textbf{VGG-22 (N)} & \textbf{Inception (N)} & \textbf{ResNet (N)} \\
\hline
\multirow{5}{*}{$\rm M_{\text{vir}}$} 
 & $14.4\mbox{-}14.6$ & B $\pm$ $\sigma$ & $0.31(8.8\%) \pm 0.42$ & $0.35(10.0\%) \pm 0.38$ & $0.38(10.8\%) \pm 0.45$ \\
 & $14.6\mbox{-}14.8$ & B $\pm$ $\sigma$ & $-0.39(-8.3\%) \pm 0.78$ & $-0.42(-8.9\%) \pm 0.74$ & $-0.27(-5.7\%) \pm 0.79$ \\
 & $14.8\mbox{-}15.0$ & B $\pm$ $\sigma$ & $-1.46(-20.0\%) \pm 1.4$ & $-1.76(-24.1\%) \pm 1.22$ & $-1.29(-17.7\%) \pm 1.4$ \\
 & $15.0\mbox{-}15.3$ & B $\pm$ $\sigma$ & $-2.32(-20.6\%) \pm 2.37$ & $-3.11(-27.5\%) \pm 2.17$ & $-2.29(-20.4\%) \pm 2.13$ \\
 & All & RMS & 1.03(20.5\%) & 1.08(21.2\%) & 1.02(20.2\%) \\
\hline
\multirow{5}{*}{$c_{\text{NFW}}$} 
 & $14.4\mbox{-}14.6$ & B $\pm$ $\sigma$ & $0.19(4.1\%) \pm 0.74$ & $0.05(1.1\%) \pm 0.79$ & $-0.01(-0.2\%) \pm 0.77$ \\
 & $14.6\mbox{-}14.8$ & B $\pm$ $\sigma$ & $0.02(0.5\%) \pm 0.68$ & $0.07(1.6\%) \pm 0.74$ & $-0.06(-1.4\%) \pm 0.71$ \\
 & $14.8\mbox{-}15.0$ & B $\pm$ $\sigma$ & $-0.17(-4.0\%) \pm 0.63$ & $0.02(0.5\%) \pm 0.69$ & $-0.07(-1.6\%) \pm 2.06$ \\
 & $15.0\mbox{-}15.3$ & B $\pm$ $\sigma$ & $-0.26(-6.3\%) \pm 0.62$ & $-0.04(-1.0\%) \pm 0.64$ & $-0.23(-5.5\%) \pm 0.64$ \\
 & All & RMS & 0.72(16.1\%) & 0.76(17.0\%) & 0.99(22.1\%) \\
\hline
\multirow{5}{*}{$c_{\text{smooth}}$} 
 & $14.4\mbox{-}14.6$ & B $\pm$ $\sigma$ & $0.16(3.3\%) \pm 0.76$ & $0.01(0.2\%) \pm 0.8$ & $-0.01(-0.2\%) \pm 0.79$ \\
 & $14.6\mbox{-}14.8$ & B $\pm$ $\sigma$ & $0.01(0.2\%) \pm 0.69$ & $0.04(0.9\%) \pm 0.74$ & $-0.06(-1.3\%) \pm 0.73$ \\
 & $14.8\mbox{-}15.0$ & B $\pm$ $\sigma$ & $-0.18(-4.0\%) \pm 0.66$ & $0.00(0.0\%) \pm 0.72$ & $-0.07(-1.5\%) \pm 1.94$ \\
 & $15.0\mbox{-}15.3$ & B $\pm$ $\sigma$ & $-0.27(-6.0\%) \pm 0.64$ & $-0.09(-2.0\%) \pm 0.63$ & $-0.23(-5.1\%) \pm 0.66$ \\
 & All & RMS & 0.73(15.3\%) & 0.77(16.2\%) & 0.97(20.4\%) \\
\hline
\multirow{5}{*}{$n_{\text{sub}}$} 
 & $14.4\mbox{-}14.6$ & B $\pm$ $\sigma$ & $-7(-1.5\%) \pm 219$ & $-9(-1.9\%) \pm 221$ & $24(5.1\%) \pm 225$ \\
 & $14.6\mbox{-}14.8$ & B $\pm$ $\sigma$ & $-64(-9.9\%) \pm 296$ & $-93(-14.4\%) \pm 295$ & $-36(-5.6\%) \pm 301$ \\
 & $14.8\mbox{-}15.0$ & B $\pm$ $\sigma$ & $-182(-18.3\%) \pm 420$ & $-253(-25.4\%) \pm 431$ & $-143(-14.4\%) \pm 447$ \\
 & $15.0\mbox{-}15.3$ & B $\pm$ $\sigma$ & $-288(-19.3\%) \pm 679$ & $-405(-27.2\%) \pm 651$ & $-260(-17.4\%) \pm 661$ \\
 & All & RMS & 306(52.2\%) & 318(54.3\%) & 309(52.7\%) \\
\hline
\multirow{5}{*}{$f_{\text{smooth}}$} 
 & $14.4\mbox{-}14.6$ & B $\pm$ $\sigma$ & $1.4(1.6\%) \pm 5.3$ & $1.1(1.3\%) \pm 5.5$ & $0.6(0.7\%) \pm 5.5$ \\
 & $14.6\mbox{-}14.8$ & B $\pm$ $\sigma$ & $0.3(0.4\%) \pm 5.6$ & $1.0(1.2\%) \pm 5.7$ & $0.0(0.0\%) \pm 5.7$ \\
 & $14.8\mbox{-}15.0$ & B $\pm$ $\sigma$ & $-0.4(-0.5\%) \pm 5.8$ & $0.8(1.0\%) \pm 5.9$ & $-0.3(-0.4\%) \pm 6.6$ \\
 & $15.0\mbox{-}15.3$ & B $\pm$ $\sigma$ & $-0.3(-0.4\%) \pm 6.5$ & $0.5(0.6\%) \pm 6.6$ & $-0.2(-0.2\%) \pm 6.7$ \\
 & All & RMS & 5.6(6.6\%) & 5.7(6.7\%) & 5.8(6.8\%) \\
\hline
\hline\hline
\end{tabular}
\tablefoot{Same as Table~\ref{tab:cnn_performance_noiseless}.}
\end{table*}

For the virial mass $M_{\rm vir}$, performance deterioration is comparable between all architectures. In particular, our models exhibit significantly inflated biases and $\sigma$ values in higher mass bins, with Inception reaching a peak bias of $-3.11 \times 10^{14}~M_\odot/h$ in the $\log_{10}(M_{\rm vir})\in[15.0, 15.3]$ range, while VGG and ResNet show slightly lower but comparable degradation. Across the entire sample, VGG and ResNet achieve the smallest RMS ($1.03$ and $1.02$ respectively), while Inception shows the highest ($1.08$), indicating higher susceptibility to overfitting noise patterns in deeper or more complex architectures. However, all our models show the tendency to underestimate $M_{\rm vir}$ with increasing mass, as in the noiseless case. 
We interpret this common noise-independent behavior as a direct reflection of the reduced exposure of CNNs to data in these extreme ranges, leading to poorer predictive accuracy and higher overall RMS than in lower-mass bins, where we find most of the data (see Figure~\ref{fig:distr}).

Estimates of NFW concentration $c_{\rm NFW}$ remain relatively robust despite noise, particularly for VGG, which preserves the lowest RMS ($0.72$) across all mass bins. 
Inception and ResNet exhibit similar performance degradation in noisy conditions, whereas VGG-Net shows more stable predictions across mass bins, with lower dispersion. This is particularly evident in the $[14.8, 15.0]$ mass range, where Resnet's higher $\sigma$ ($\sim2.0$) reveals greater sensitivity to outlier predictions compared to VGG's more robust behavior.
A similar behavior is observed for the smoothed concentration $c_{\rm{smooth}}$, with VGG again producing the lowest RMS (0.73).

Substructure-related parameters are more severely impacted by noise. The number of subhalos $n_{\rm{sub}}$ shows markedly high RMS values for all models, coupled with a strong bias (and an evenly stronger dispersion), reflecting the models’ tendency to wrongly predict subhalo counts under noisy conditions. This suggests a failure to discriminate between noise-induced fluctuations and real small-scale substructures, due to the loss of small-scale information introduced by noise. As a consequence, on average, CNNs are substantially underpredicting the number of subhalos, especially in high-mass clusters. VGG and Inception fare slightly better than ResNet but still exhibit substantial errors, reinforcing the difficulty of recovering discrete subhalo properties in noisy conditions. In contrast, the smooth component mass fraction $f_{\rm{smooth}}$ remains a more inferable target: although noise increases overall RMS values (up to $0.058$ for ResNet), biases remain relatively low across models ($\sim 1\%$ or less), indicating that global substructure content is still partially recoverable (as $f_{\rm{smooth}}=1-f_{\rm sub}$), albeit with reduced precision as the mean dispersion values increase up to $\sim 6.7 \%$.

In general, VGG demonstrates the best robustness to noise, particularly in low-mass regimes and for global properties such as $c_{\rm NFW}$ and $f_{\rm smooth}$. The performance degradation of Inception and ResNet, especially at high masses, suggests that their more complex representations may be more vulnerable to overfitting noise features.

\begin{figure}
    \centering
    {\includegraphics[width=0.5\textwidth]{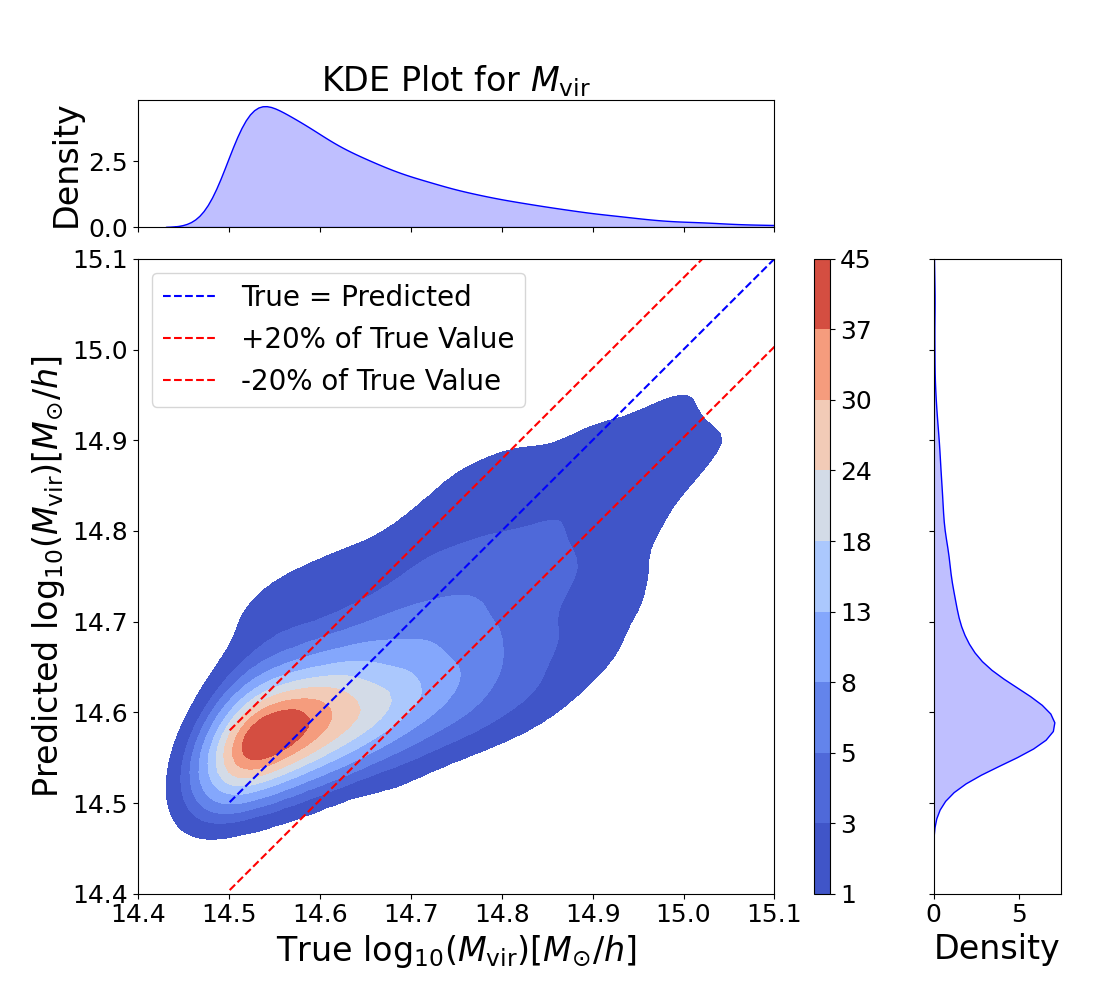}}\hfill
    {\includegraphics[width=0.5\textwidth]{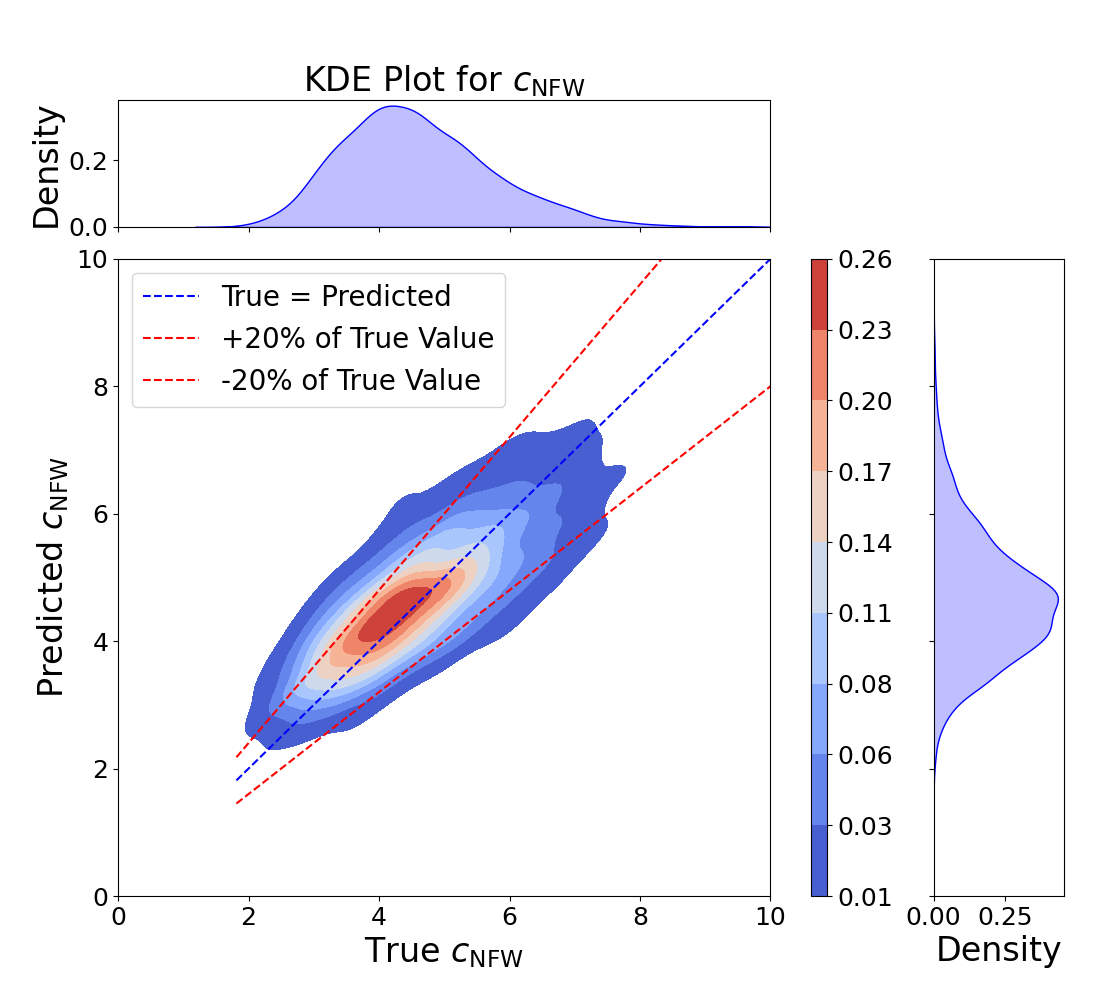}}\\
    \caption{KDE plots showing predicted vs target values for $M_{\rm vir}$ (upper panel) and $c_{\rm NFW}$ (lower panel) obtained applying our VGG-22 model on noisy reduced shear maps of clusters located at $z = 0.25$.}
    \label{fig:kdeplotvggnoise_m_c}
\end{figure}

\begin{figure*}
    \centering
    \includegraphics[width = 1\textwidth]{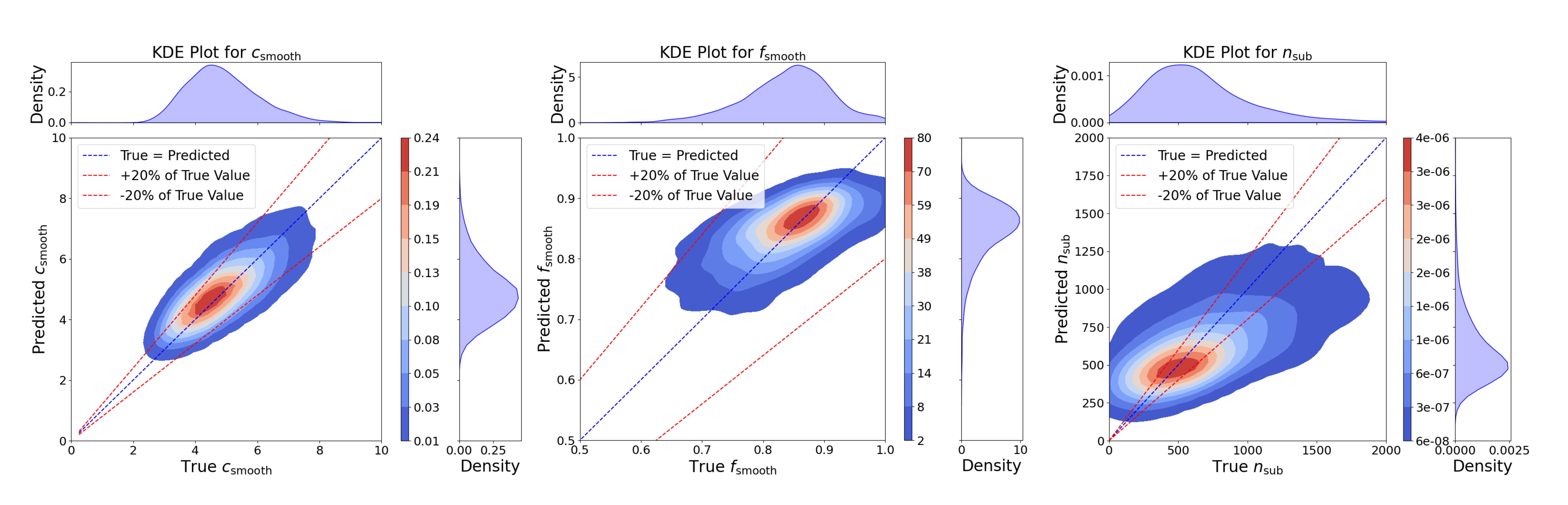}
    
    \caption{Same as Figure~\ref{fig:kdeplotvggnoise_m_c} but for $c_{\rm smooth}$ (left panel), $f_{\rm smooth}$ (center panel)  and $n_{\rm sub}$ (right panel).}
    \label{fig:kdeplotsvggnoisepanel}
\end{figure*}

Figure~\ref{fig:kdeplotvggnoise_m_c} and Figure~\ref{fig:kdeplotsvggnoisepanel} show 2D kernel density estimates (KDEs) of predicted versus true values for key parameters as inferred by the VGG-22 model.

The predictions for $ M_{\rm{vir}}$ show a mean systematic overestimation at lower masses, but present an increasing dispersion and a tendency to underestimate real values at the high-mass end. This behavior reflects the bias and $\sigma$ trends reported in Table~\ref{tab:cnn_performance_noisy}, confirming that noisy maps challenge the model’s ability to precisely recover $ M_{\rm{vir}}$ in massive systems.

For both $c_{\rm{NFW}}$ and $c_{\rm{smooth}}$, the KDE contours are tightly centered along the identity line, with narrow scatter, indicating that these shape-related features are more robust to noise and are effectively captured by the VGG architecture.

The smooth component mass fraction $f_{\rm{smooth}}$ is the best-predicted parameter under noise conditions, with predictions clustering tightly around the one-to-one line and staying well within the $\pm20\%$ bands across the entire range. This confirms that global properties of substructures are learnable and are less affected by small-scale noise.

In contrast, the number of subhalos $n_{\rm{sub}}$ is the least accurately predicted quantity. The KDE is broadly spread, with a visible under-prediction trend at high subhalo counts and a large dispersion, especially at higher $n_{\rm{sub}}$. This supports the interpretation that subhalo counts are more sensitive to stochastic variations and noise because of insufficient resolution in the input maps.

In conclusion, while noise limits model sensitivity to fine structural details, the networks retain reliable predictive power across all clusters' large-scale parameters, except $n_{\rm{sub}}$.

\subsection{Tangential shear profile fitting vs. CNN-based predictions}

In this section, we present a comparative analysis of the results obtained on noisy maps by our DL-based algorithms and by traditional reduced-tangential shear-profile-fitting methods for the estimation of cluster parameters.

For the classical approach, we employ the Python library \texttt{pyLensLib} \citep{meneghetti2021introduction}, which provides tools to analytically model the lensing properties of pseudo-elliptical NFW halos. Using the \texttt{nfwell} class, one could generate convergence and shear maps for any cluster by inputting cosmological model parameters, angular diameter distances, and halo properties (i.e., mass $M_{\rm vir}$, concentration $c_{\rm vir}$, axis ratio $q$, and position angle $\alpha$).

\begin{figure*}
    \centering
    {\includegraphics[width=0.33\textwidth]{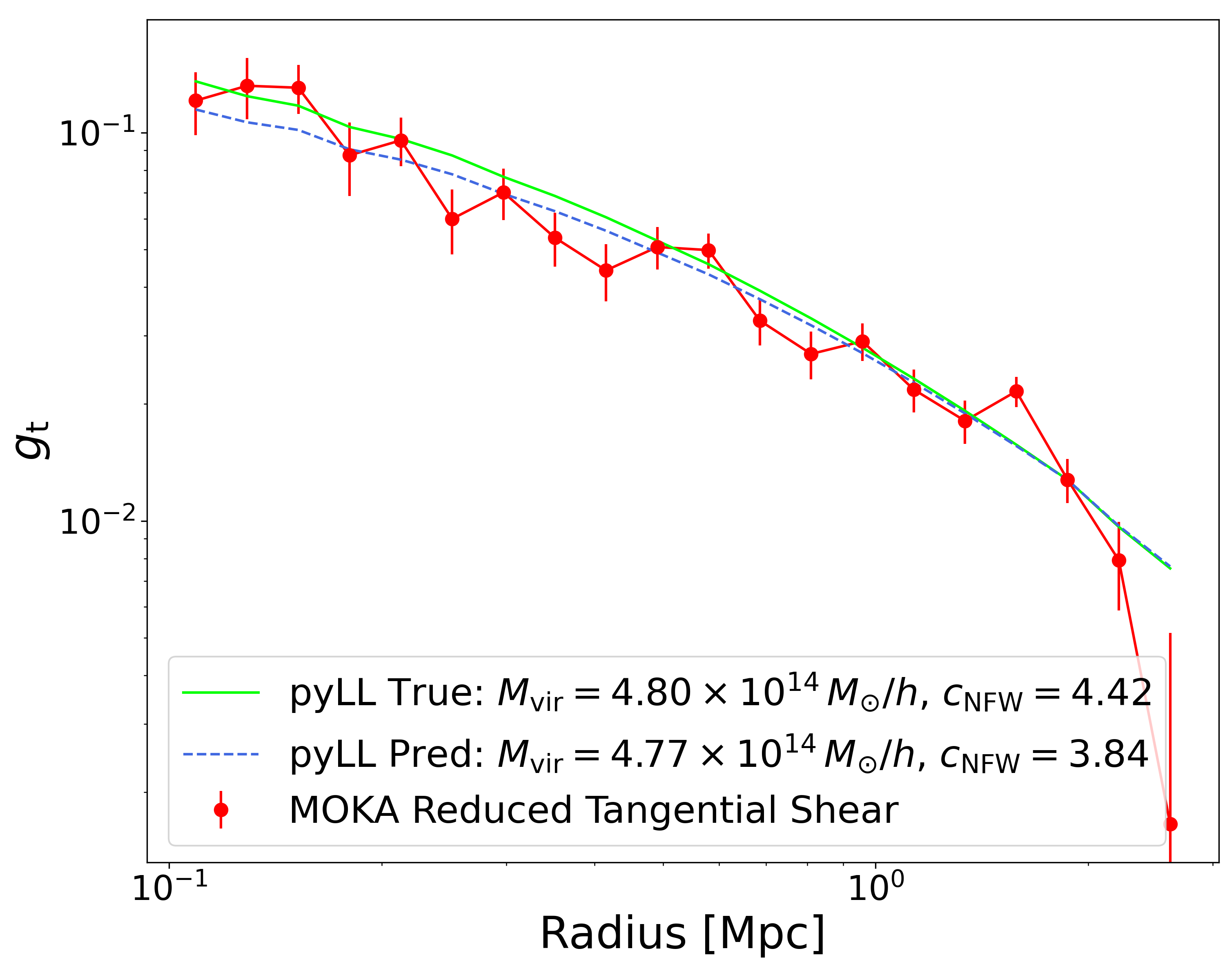}}\hfill
    {\includegraphics[width=0.33\textwidth]{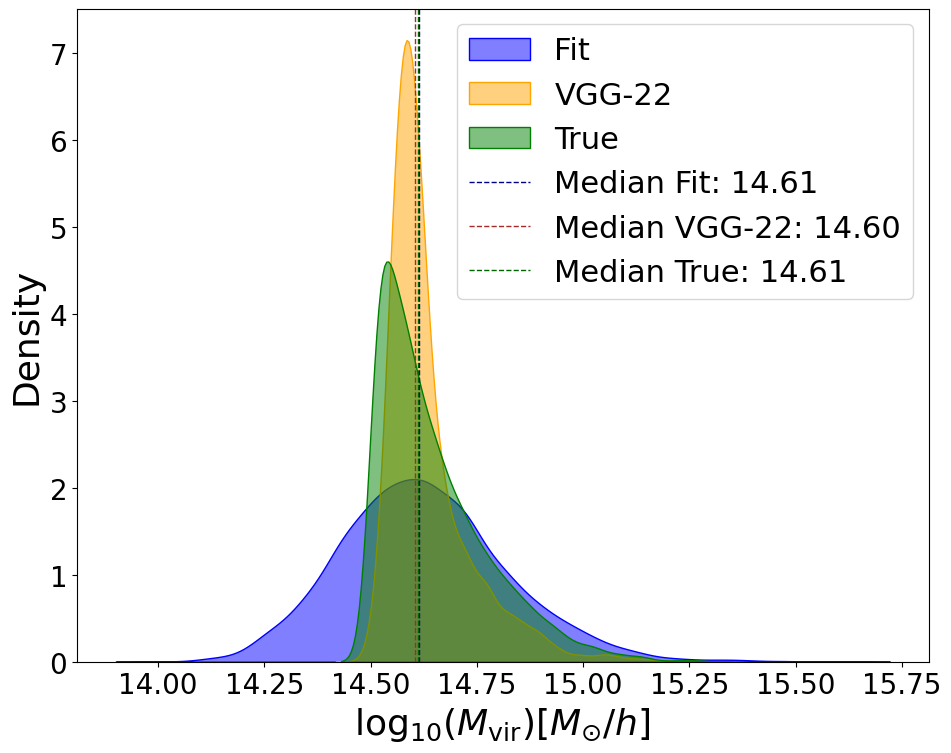}}\hfill
    {\includegraphics[width=0.33\textwidth]{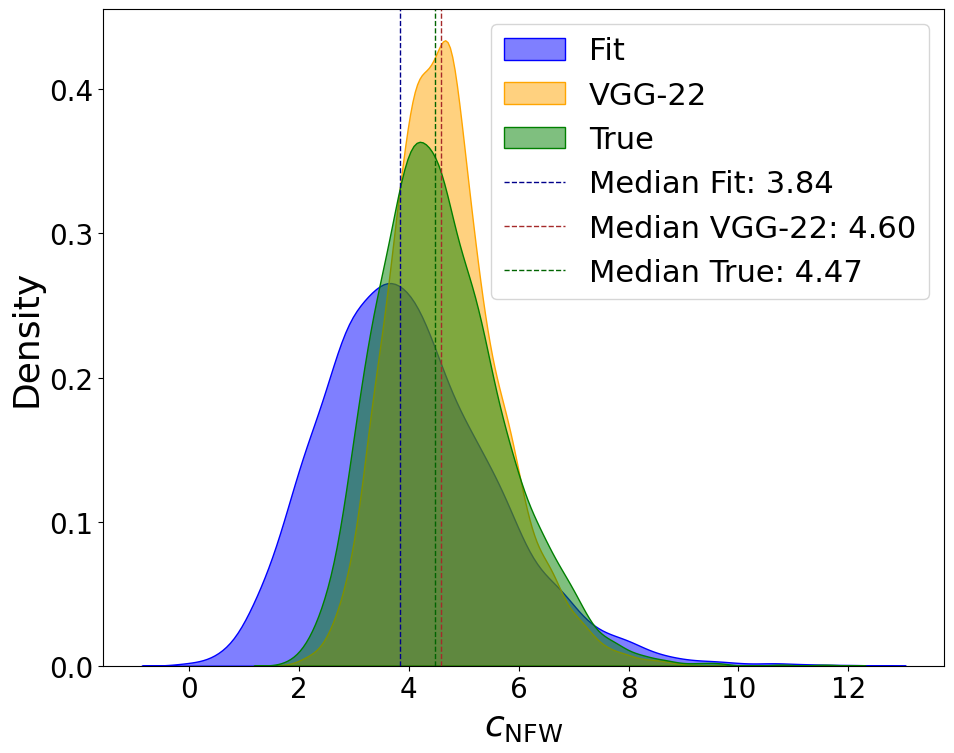}}\\
    \caption{Left panel: example of reduced tangential shear profile fitting for a single galaxy cluster. Red points with error bars show the \texttt{MOKA}-generated reduced tangential shear profile ($g_t$) as a function of radius. The green line represents the theoretical NFW profile created with the \texttt{pyLensLib} library using the true cluster parameters ($M_{\rm vir} = 4.80 \times 10^{14}$ $M_\odot/h$; $c_{\rm NFW} = 4.42$), while the blue line shows the equivalent of the best-fit NFW model from traditional profile fitting ($M_{\rm vir} = 4.77 \times 10^{14}$ $M_\odot/h$; $c_{\rm NFW} = 3.84$). Center and right panels: comparison of mass and concentration parameter distributions between traditional fitting methods (blue), CNN predictions (orange), and true values (green). Vertical dashed lines indicate the median values for each method.}
    \label{fig:fitpanel}
\end{figure*}

For each \texttt{MOKA}-simulated cluster, we extract the tangential reduced shear profile (red dotted line, see Figure~\ref{fig:fitpanel}a), excluding the inner region ($r < 0.1 ~{\rm Mpc}$) where strong lensing effects invalidate the weak lensing approximation. Finally, we fit the extracted profile with an NFW model using the \texttt{nfwell} class to obtain the best-fit parameters for physical mass $M_{\rm vir}$ and concentration $c_{\rm NFW}$.

Figure~\ref{fig:fitpanel} shows the results of the fitted predictions for mass (Figure~\ref{fig:fitpanel}b) and NFW concentration (Figure~\ref{fig:fitpanel}c) compared to the predictions of the VGG-22 model (orange distributions in Figure~\ref{fig:fitpanel}), obtained on the noisy weak-lensing maps of the test set. Ground truth values (green distributions in Figure~\ref{fig:fitpanel}) served as a benchmark for this comparison.

For cluster mass estimation, both methods demonstrated high accuracy, with median values of $\log_{10}(M_{\rm vir})[M_\odot/h] = 14.61$ for the traditional fit, matching the true median of $14.61$, and $14.60$ for VGG-22, showing a negligible underestimation of $-0.01$ dex ($\sim-2.3\%$ in linear units). This precision suggests that mass estimation is, on average, robust to the choice of methodology when applied to noisy data. 

The concentration estimates reveal more pronounced differences. The traditional fit underestimates $c_{\rm NFW}$ by $\sim 14\%$ (median $c_{\rm NFW} = 3.84$ versus true $4.47$), likely due to noise amplifying small-scale deviations from the idealized NFW profile and suppressing the measured curvature of the inner density profile. In contrast, VGG-22 achieves better agreement with the ground truth (median $c_{\rm NFW} = 4.60$, corresponding to a median overestimation of $+2.9\%$ ). This behavior suggests greater robustness to noise-induced perturbations in the shear field, indicating that VGG-Net's feature extraction preserves more of the central concentration signal despite noise. This advantage arises from the network's capacity to learn triaxial features in the data that are not fully captured by the radial profile fitting. Cluster triaxiality is a primary driver of scatter and systematic error in traditional lensing studies, creating a significant orientation bias when deprojecting 2D data into 3D estimates \citep{meneghetti10b,giocoli12a}. This deprojection typically assumes spherical symmetry, which causes the 3D mass and concentration to be overestimated when the major axis is aligned with the line of sight and underestimated when it is perpendicular. 

Our results show that this orientation bias is less important for CNNs because they can learn to better characterize the 3D orientation of the halo by extracting features from the full 2D shear field rather than relying on a radially averaged profile. Because the network can account for these triaxial features, its concentration estimates are significantly less biased than those derived from traditional fitting methods.

Overall, the CNN demonstrated superior precision, with tighter distributions for both mass and concentration compared to the traditional fit. 
The traditional method, while being more physically interpretable, showed higher variance in concentration estimates, reflecting its sensitivity to choices such as radial binning and the exclusion of the strong-lensing region ($r < 0.1$ Mpc).

The traditional fitting approach remains valuable for its interpretability, as it directly links measurements to the underlying NFW model. However, this method requires careful handling of systematics, including noise suppression and masking of the inner cluster region where the weak-lensing approximation breaks down. These steps introduce subjectivity and computational cost, especially for large surveys.

CNNs, by contrast, process the full shear map without explicit masking or binning, enabling faster and more automated analysis. This advantage comes at the cost of interpretability, as the network's decision-making process is not inferable. However, mass underestimation trends observed in CNN predictions underscore the need for rigorous validation against simulations to identify and correct such biases. 

\section{\label{sc:concl} Conclusions }

In this work, we investigated the efficacy of Deep CNNs for estimating the physical parameters of galaxy clusters from simulated weak lensing shear maps. We developed and evaluated several architectures, including VGG, Inception-v4, and Inception-ResNet-v2, and assessed their performance under both idealized, noiseless conditions and more realistic scenarios incorporating simulated Euclid-like shape noise. Our analysis demonstrates that DL offers a powerful and robust alternative to traditional methods for cluster cosmology.

Our primary finding is that CNNs exhibit remarkable robustness to significant noise, establishing them as a valuable tool for future large-scale surveys. Even when noise levels were sufficient to obscure fine-scale substructures, our models, particularly the VGG-based architectures, reliably recovered large-scale halo properties such as virial mass ($M_{\rm vir}$) and concentration ($c_{\rm NFW}$, $c_{\rm smooth}$). Notably, these results were achieved without any pre-processing or smoothing of the input shear maps, highlighting the networks' intrinsic ability to learn and filter relevant features directly from raw, noisy data. While the recovery of discrete subhalo counts ($n_{\rm sub}$) was severely degraded by noise, the smooth component mass fraction ($f_{\rm smooth}$) remained a learnable parameter, suggesting that global properties are more resilient to noise than localized, small-scale features.

A comparative analysis against traditional tangential shear profile fitting further underscored the advantages of our DL approach. Specifically, classical parametric modeling typically estimates cluster mass by fitting the azimuthally averaged tangential reduced shear profile; this 1D reduction inevitably collapses the non-Gaussian and aspherical information present in the full shear field.

While both methods accurately estimated cluster mass, our VGG-22 model demonstrated superior performance in estimating concentration, yielding results with significantly lower bias and reduced scatter compared to the fitting method. This superiority stems from the CNN's ability to leverage the full 2D information of the shear field. 

Looking forward, our results suggest several paths for future work. The current study was based on dark-matter-only simulations. To enhance the realism of our training data and ensure the models' applicability to real observational data, the next logical step is to incorporate baryonic physics. Future work should focus on training and validating these networks using state-of-the-art hydrodynamical simulations that model the complex interplay between dark matter, gas, and stars. This will allow us to quantify the impact of baryonic effects on CNN-based parameter inference and to develop models that are robust to these additional physical processes.

In conclusion, our work confirms that CNNs are not only a viable but also a highly competitive method for estimating galaxy cluster parameters from weak lensing data. Their ability to extract information from noisy, high-dimensional data surpasses that of traditional techniques in key areas, paving the way for faster, more automated, and more precise analysis of the vast datasets expected from upcoming surveys like Euclid, LSST and the Vera C. Rubin Observatory.
%
%

\begin{acknowledgements}
M.M. was supported by INAF Grants “The Big-Data era of cluster lensing" and "Probing Dark Matter and Galaxy Formation in Galaxy Clusters through Strong Gravitational Lensing", and ASI Grant n. 2024-10-HH.0 "Attività scientifiche per la missione Euclid – fase E". The research activities described in this paper have been co-funded by the European Union – NextGeneration EU within PRIN 2022
project no. 20229YBSAN – Globular clusters in cosmological simulations and in lensed fields: from their birth to the present epoch.

L.M. acknowledges the financial contribution from the PRIN-MUR 2022 20227RNLY3 grant “The concordance cosmological model: stress-tests with galaxy clusters” supported by Next Generation EU and from the ASI grant n. 2024-10-HH.0 “Attività scientifiche per la missione Euclid – fase E”
\end{acknowledgements}

%
%

\bibliography{Euclid, Q1, z_biblio,globalbibs} 

%

\begin{appendix}

\section{\label{App:lensing_maps} Convergence and Shear Map Generation}

The computation of weak lensing observables from the mass distribution follows the standard approach based on the gravitational lensing potential and its derivatives. This section describes the numerical implementation used to generate the convergence ($\kappa$) and shear ($\gamma_1$, $\gamma_2$) maps from the mass distributions produced by the \texttt{MOKA} software.

\subsection{Mathematical Framework}

The weak gravitational lensing formalism relates the convergence to the 2D gravitational potential
$\phi(\mathbf{x})$ through the following equation:

\begin{equation}
\kappa(\mathbf{x}) = \frac{1}{2}\nabla^2 \phi(\mathbf{x}),
\label{eq:convergence}
\end{equation}
where $\kappa$ is the convergence field. The shear components $\gamma_1$ and $\gamma_2$ are instead obtained as in Eq.~\ref{eq:shear1} and Eq.~\ref{eq:shear2}.

\subsection{Fourier Space Implementation}

Given the computational efficiency and the nature of the differential operators involved, all calculations are performed in Fourier space using Fast Fourier Transform (FFT) techniques implemented through the FFTW3 library \citep{frigoFFTW}.

\paragraph{Zero-padding and Fourier Transform}
To minimize boundary effects, the input convergence maps are zero-padded by a factor $z$ (typically 2 or 4), expanding the original $N \times N$ pixel map to $(zN) \times (zN)$. The convergence field is then transformed to Fourier space:

\begin{equation}
\tilde{\kappa}(\mathbf{k}) = \mathcal{F}[\kappa(\mathbf{x})]
\end{equation}

where $\mathbf{k} = (k_x, k_y)$ are the Fourier modes defined as:

\begin{equation}
k_x = \frac{2\pi i}{L \cdot z}, \quad k_y = \frac{2\pi j}{L \cdot z}
\end{equation}

with $i, j$ being the pixel indices, $L$ the physical box size in Mpc/h, and $z$ the zero-padding factor.

\paragraph{Gravitational Potential Calculation}
The lensing potential in Fourier space is obtained by inverting the Poisson equation (Eq.~\ref{eq:convergence}):

\begin{equation}
\tilde{\phi}(\mathbf{k}) = -\frac{2\tilde{\kappa}(\mathbf{k})}{|\mathbf{k}|^2}
\label{eq:potential_fourier}
\end{equation}

where $|\mathbf{k}|^2 = k_x^2 + k_y^2$. The zero mode ($\mathbf{k} = 0$) is set to zero to avoid divergences, which corresponds to fixing the arbitrary constant in the potential.

\paragraph{Shear Components in Fourier Space}
The shear components are computed using the second derivatives of the potential. In Fourier space, spatial derivatives become simple multiplications by the corresponding Fourier modes:

\begin{equation}
\tilde{\gamma}_1(\mathbf{k}) = \frac{1}{2}(k_x^2 - k_y^2)\tilde{\phi}(\mathbf{k})
\label{eq:gamma1_fourier}
\end{equation}

\begin{equation}
\tilde{\gamma}_2(\mathbf{k}) = k_x k_y \tilde{\phi}(\mathbf{k})
\label{eq:gamma2_fourier}
\end{equation}

\paragraph{Optional Fourier Filtering}
To reduce high-frequency noise and potential numerical artifacts, an optional Gaussian filter can be applied in Fourier space:

\begin{equation}
\tilde{\kappa}_{\text{filtered}}(\mathbf{k}) = \tilde{\kappa}(\mathbf{k}) \exp\left(-\frac{|\mathbf{k}|^2}{2\sigma_g^2}\right)
\label{eq:gaussian_filter}
\end{equation}

where $\sigma_g = (2\pi/L/z) \cdot (N_{\text{pix}}/N_{\text{filter}})$ and $N_{\text{filter}}$ is the smoothing scale in pixels.

\paragraph{Inverse Fourier Transform}
Finally, the shear components are transformed back to real space through inverse FFT:

\begin{equation}
\gamma_1(\mathbf{x}) = \mathcal{F}^{-1}[\tilde{\gamma}_1(\mathbf{k})]
\end{equation}

\begin{equation}
\gamma_2(\mathbf{x}) = \mathcal{F}^{-1}[\tilde{\gamma}_2(\mathbf{k})]
\end{equation}

\subsubsection{Computational Implementation}

The algorithm is implemented in the \texttt{fftw\_lens} class, which handles both square and rectangular maps. The main computational steps are executed in the \texttt{lensingComponents()} method with the following steps:

\begin{enumerate}
\item The input convergence map is zero-padded and centered within the expanded grid.
\item The padded convergence map is transformed to Fourier space using a 2D real-to-complex FFT.
\item The potential and shear components are computed according to Eq.~(\ref{eq:potential_fourier}), (\ref{eq:gamma1_fourier}), and (\ref{eq:gamma2_fourier}).
\item The inverse FFT for each lensing quantity is performed only when explicitly requested through the corresponding accessor methods (\texttt{shear1()}, \texttt{shear2()}).
\end{enumerate}

This implementation provides an efficient and numerically stable method for computing weak lensing observables, suitable for large-scale analysis of thousands of simulated cluster maps required for training DL models.

\subsection{Additional Lensing Quantities}

Beyond the primary shear components, the implementation also computes additional lensing quantities:

\begin{itemize}
\item Deflection angles: $\alpha_1 = -\partial\phi/\partial x$ and $\alpha_2 = -\partial\phi/\partial y$
\item Flexion terms: Higher-order lensing effects $F_1$, $F_2$, $G_1$, $G_2$ computed directly from the convergence field and potential derivatives
\end{itemize}

These quantities, while not used in the current analysis, provide a complete characterization of the gravitational lensing field and may be valuable for future extensions of this work.

\section{\label{sc:app A} VGG-Net Architecture}
\label{sc:app:vgg_arch}

This appendix details the specific architectural modifications applied to the standard VGG-Net framework to adapt it for the regression task of predicting galaxy cluster halo parameters from weak lensing maps. While our implementation is inspired by the original D configuration, commonly referred to as ``VGG-16'' due to its composition of 16 convolutional layers \citep{simonyan2015deep}, several key changes were necessary to suit our specific scientific goal.

A primary adaptation was to accommodate our two-channel input data (shear components $\gamma_1$ and $\gamma_2$) instead of the standard three-channel RGB images. Furthermore, to improve training stability and accelerate convergence, each $3\times3$ convolutional layer is followed by a Batch Normalization layer \citep{ioffe2015batch}. This layer stabilizes training by normalizing layer inputs to zero mean and unit variance, mitigating internal covariate shift. Finally, the output is passed to a Leaky-ReLU activation function with a negative slope $a=0.3$. This function, which calculates the output value for each neuron, introduces a small slope $a$ for negative values, allowing a non-zero gradient for negative inputs and preventing the neurons from becoming inactive and halting learning during training \citep{pedamonti2018comparison}.
A detailed example of the convolutional layer used in the VGG and Inception models is shown in Table~\ref{tabconv2d}. 

\begin{table}[h]
\centering
  \captionsetup{width=0.8\linewidth}
  \caption{Scheme of the basic convolutional block employed in the VGG-Net, Inception-v4, and Inception-ResNet-v2 models. \texttt{Conv2D}, \texttt{BatchNorm2d} and \texttt{LeakyReLU} are commands imported in Python from the Pytorch library. In the VGG-Net models, the $kernel\_size$ is fixed at $3\times3$, with the stride and padding both set to 1. Conversely, for the Inception architectures, the \texttt{Conv2D} settings are adjusted according to the authors' recommendations outlined in the original papers.}
  \begin{tabular}{|c|}
  \hline
  \rule{0pt}{12pt}\texttt{Conv2D} \\
  \hline
  \rule{0pt}{12pt}\texttt{BatchNorm2d}($out\_channels$, $eps=10^{-5}$, $momentum=0.1$) \\
  \hline
  \rule{0pt}{12pt}\texttt{LeakyReLU}($negative\_slope=0.3$, $inplace=True$) \\
  \hline
\end{tabular}
\label{tabconv2d}
\end{table}

The most significant modification was converting the network from a classifier to a regressor. The original VGG-16 architecture terminates with a softmax activation function to produce class probabilities. We replaced this classification head with a new sequence of Fully Connected (FC) layers designed for regression. This new head maps the flattened feature vector from the convolutional base to a final output vector of five continuous parameters. To mitigate overfitting in these dense layers, we incorporated dropout modules \citep{JMLR:v15:srivastava14a} between the first three FC blocks. Dropout modules temporarily remove certain units from the network, including their incoming and outgoing connections. The units selected for removal are chosen randomly, each with an independent fixed probability denoted as $p$. In our case, we set $p=0.3$, indicating that 30\% of the following connections are randomly dropped. At test time, a single model without dropout is used: the weights of this network are scaled-down versions of the trained weights. In particular, if a unit is retained with probability $p$ during training, the outgoing weights of that unit are multiplied by $p$ at the test time. 

In this work, we explored several variants of this adapted architecture. Table~\ref{tab:vgg_variants} provides a comparative overview of the original VGG-16 configuration alongside two of our key implemented models: VGG-19 and VGG-22. These models primarily differ in the number of channels in their convolutional blocks, which directly influence the network's capacity and total number of trainable parameters, which are $\sim 69$ million and $\sim 85$ million for VGG-19 and VGG-22, respectively. The FC layers were also adjusted accordingly to handle the output from their respective convolutional bases.

\begin{table}[h!]
  \centering
  \captionsetup{width=0.9\linewidth}
  \caption{Comparison of VGG-inspired architectures. The table details the original VGG-16 configuration and two custom models implemented in this work. The convolutional layer format is “conv$\langle$receptive field size$\rangle$-$\langle$number of channels$\rangle$”. For brevity, Batch Normalization and Leaky-ReLU layers, which follow each convolution, are not shown.}
  \begin{tabular}{|c|c|c|}
    \hline
    \rule{0pt}{10pt}\textbf{VGG-16} & \rule{0pt}{10pt}\textbf{VGG-19} & \rule{0pt}{10pt}\textbf{VGG-22} \\
    \hline
    \multicolumn{3}{|c|}{\rule{0pt}{10pt}input ($224\times224$ 3D/2D image)} \\
    \hline
    \multirow{2}{*}{\shortstack{conv3-64 \\ conv3-64}} & \multirow{2}{*}{\shortstack{conv3-64 \\ conv3-64}} & \multirow{2}{*}{\shortstack{conv3-32 \\ conv3-32}} \\
    & & \\
   \hline
   \multicolumn{3}{|c|}{maxpool} \\
   \hline
    \multirow{2}{*}{\shortstack{conv3-128 \\ conv3-128}} & \multirow{2}{*}{\shortstack{conv3-128 \\ conv3-128}} & \multirow{2}{*}{\shortstack{conv3-64 \\ conv3-64}} \\
    & & \\
    \hline
    \multicolumn{3}{|c|}{maxpool} \\
    \hline
   \multirow{3}{*}{\shortstack{conv3-256 \\ conv3-256 \\ conv3-256}} & \multirow{3}{*}{\shortstack{conv3-256 \\ conv3-256 \\ conv3-256}} & \multirow{3}{*}{\shortstack{conv3-128 \\ conv3-128 \\ conv3-128}} \\
    & & \\
    & & \\
   \hline
    \multicolumn{3}{|c|}{maxpool} \\
    \hline
   \multirow{3}{*}{\shortstack{conv3-512 \\ conv3-512 \\ conv3-512}} & \multirow{3}{*}{\shortstack{conv3-512 \\ conv3-512 \\ conv3-512}} & \multirow{3}{*}{\shortstack{conv3-256 \\ conv3-256 \\ conv3-256}} \\
    & & \\
    & & \\
  \hline
   \multicolumn{3}{|c|}{maxpool} \\
    \hline
    \multirow{7}{*}{\shortstack{conv3-512 \\ conv3-512 \\ conv3-512}} & \multirow{7}{*}{\shortstack{conv3-512 \\ conv3-512 \\ conv3-512}} & \rule{0pt}{10pt}conv3-512 \\
    & & conv3-512 \\
    & & conv3-512 \\
    \cline{3-3} 
    & & maxpool \\
    \cline{3-3}
    & & \rule{0pt}{10pt}conv3-512 \\
    & & conv3-512 \\
    & & conv3-512 \\
   \hline
   \multicolumn{3}{|c|}{maxpool} \\
    \hline
    \multirow{6}{*}{\shortstack{FC-4096 \\ FC-4096}} & \multirow{6}{*}{\shortstack{FC-2048 \\ FC-1024 \\ FC-512 \\ FC-256 \\ FC-128}} & \multirow{6}{*}{\shortstack{FC-2048 \\ FC-1024 \\ FC-512 \\ FC-256 \\ FC-128}} \\
    & & \\
    & & \\
    & & \\
    & & \\
    & & \\
    \hline
    FC-1000 & FC-5 & FC-5 \\
    \hline
  \end{tabular}
  \label{tab:vgg_variants}
\end{table}

\section{Inception and ResNet Architectures}
\label{sc:app:inception}

This appendix provides a detailed account of the Inception-v4 and Inception-ResNet-v2 architectures implemented in this work. Building upon the foundational concepts introduced in Sect.~\ref{sc:incnetowrk} and~\ref{sc:resnet}, we elaborate on the theoretical motivations, structural components, and specific adaptations made to tailor these models for our regression task.

\subsection{Architectural Principles and Core Components}

The Inception family of networks marked a significant departure from the simple sequential stacking of convolutional layers seen in architectures like VGG-Net. The core design concept is to perform multi-scale processing in parallel within a single network module, allowing the model to capture features at various levels of abstraction simultaneously.

\begin{figure}[h!]
    \centering
    \subfloat[Inception-v4 A block]{%
        \includegraphics[width=0.48\textwidth]{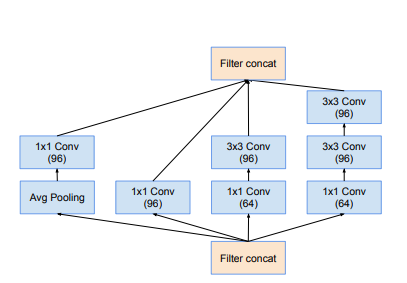}%
        \label{fig:inc_block_a}
    }
    \hfill
    \subfloat[Inception-ResNet-v2 A block]{%
        \includegraphics[width=0.48\textwidth]{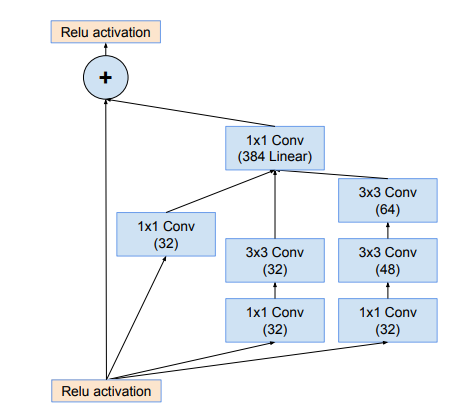}
        \label{fig:inc_res_block_a}
    }
    \caption{Architectural comparison of the fundamental "A" blocks for Inception-v4 and Inception-ResNet-v2. Panel (a) shows the parallel, multi-scale convolutional pathways of the standard Inception module. Panel (b) illustrates the integration of a residual connection, where the input is added to the output of the parallel filter bank. Figures adapted from \cite{szegedy2016inceptionv4}.}
    \label{fig:inc_blocks}
\end{figure}

\paragraph{Factorized Convolutions}
A key innovation for improving computational efficiency is the use of factorized convolutions. Instead of employing large, computationally expensive filters (e.g., $5\times5$ or $7\times7$), Inception modules decompose these operations into a sequence of smaller filters. For instance, a $5\times5$ convolution is replaced by two stacked $3\times3$ convolutions (see Figure~\ref{fig:inc_block_a}). This preserves the receptive field size while significantly reducing the number of parameters. Further optimization is achieved by decomposing an $n \times n$ convolution into a sequence of $1 \times n$ and $n \times 1$ convolutions, which is particularly effective for larger filter sizes.

\paragraph{Grid Reduction Modules}
A common challenge in deep networks is downsampling the feature map grid without creating a "representational bottleneck", i.e., an abrupt loss of information. Inception architectures address this with specialized reduction modules. These modules employ a parallel structure, combining a traditional max-pooling branch with a strided convolutional branch. By processing the input in parallel and concatenating the results, the network can reduce the grid dimensions (e.g., from $35\times35$ to $17\times17$) while preserving a rich feature representation.

\paragraph{Residual Connections}
The Inception-ResNet variants incorporate the central idea from Residual Networks (ResNets) \citep{he2015deep}. By adding shortcut or "skip" connections that bypass one or more layers, the network can learn a residual mapping (see Sect.~\ref{sc:resnet}). The output of a block is the sum of its input and the transformations learned by its layers. This mechanism facilitates gradient flow through very deep networks, mitigating the vanishing gradient problem and enabling the training of significantly deeper models than was previously feasible. The hybrid Inception-ResNet architecture applies this principle by adding a shortcut connection around the parallel convolutional filters of an Inception module, as illustrated in Figure~\ref{fig:inc_res_block_a}.

\subsection{Implementation and Adaptations}

Our implementation of the Inception architectures involved several modifications to the original designs to optimize them for predicting halo parameters from weak lensing maps.

\paragraph{Base Convolutional Block}
All convolutional operations in our models are built upon a fundamental block consisting of three sequential components from the PyTorch library: a 2D convolution (\texttt{Conv2D}), a batch normalization layer (\texttt{BatchNorm2d}), and a Leaky-ReLU activation function (\texttt{LeakyReLU}). The specific parameters for the convolution (kernel size, stride, padding) vary according to the definitions of each Inception module, while the batch normalization and activation parameters remain consistent, as detailed in Table~\ref{tabconv2d}.

\paragraph{Model Variants and Structure}
To process our native $512 \times 512$ pixel weak lensing maps, which are larger than the original $299 \times 299$ inputs for which the networks were designed, we modified the stem of the Inception-v4 architecture. As detailed in Table~\ref {tab:stem_inc}, this was achieved by changing the padding of several convolutional and pooling layers from \texttt{0} (valid padding) to \texttt{1} (same padding), indicated by \texttt{0[1]} in the table. This adjustment ensures the stem's output feature map has the correct dimensions to be processed by the subsequent Inception blocks. In contrast, the Inception-ResNet-v2 stem, shown in Table~\ref {tab:stem_res}, did not require modification as its combination of padded and unpadded convolutions inherently handles larger input dimensions without producing incompatible output sizes.

\renewcommand{\arraystretch}{1.2}
\begin{table*}[h]
\centering
\caption{Operations in the Inception-v4 Stem (\texttt{StemInc})}
\label{tab:stem_inc}
\begin{tabularx}{\textwidth}{l l c c c c c}
\hline\hline
\textbf{Layer Name} & \textbf{Operation} & \textbf{In Channels} & \textbf{Out Channels} & \textbf{Kernel} & \textbf{Stride} & \textbf{Padding} \\
\hline
\texttt{conv2d\_1a\_3x3}       & Conv2d      & \texttt{in\_channels} & 32  & 3$\times$3 & 2 & 0[1] \\
\texttt{conv2d\_2a\_3x3}       & Conv2d      & 32  & 32  & 3$\times$3 & 1 & 0[1] \\
\texttt{conv2d\_2b\_3x3}       & Conv2d      & 32  & 64  & 3$\times$3 & 1 & 1 \\
\texttt{mixed\_3a\_branch\_0}  & MaxPool2d   & 64  & 64  & 3$\times$3 & 2 & 0[1] \\
\texttt{mixed\_3a\_branch\_1}  & Conv2d      & 64  & 96  & 3$\times$3 & 2 & 0[1] \\
\texttt{concat\_3a}            & Concat      & 64+96 & 160 & — & — & — \\
\texttt{mixed\_4a\_branch\_0\_conv1} & Conv2d & 160 & 64  & 1$\times$1 & 1 & 0 \\
\texttt{mixed\_4a\_branch\_0\_conv2} & Conv2d & 64  & 96  & 3$\times$3 & 1 & 0[1] \\
\texttt{mixed\_4a\_branch\_1\_conv1} & Conv2d & 160 & 64  & 1$\times$1 & 1 & 0 \\
\texttt{mixed\_4a\_branch\_1\_conv2} & Conv2d & 64  & 64  & 1$\times$7 & 1 & (0, 3) \\
\texttt{mixed\_4a\_branch\_1\_conv3} & Conv2d & 64  & 64  & 7$\times$1 & 1 & (3, 0) \\
\texttt{mixed\_4a\_branch\_1\_conv4} & Conv2d & 64  & 96  & 3$\times$3 & 1 & 0[1] \\
\texttt{concat\_4a}            & Concat      & 96+96 & 192 & — & — & — \\
\texttt{mixed\_5a\_branch\_0}  & Conv2d      & 192 & 192 & 3$\times$3 & 2 & 0[1] \\
\texttt{mixed\_5a\_branch\_1}  & MaxPool2d   & 192 & 192 & 3$\times$3 & 2 & 0[1] \\
\texttt{concat\_5a}            & Concat      & 192+192 & 384 & — & — & — \\
\hline\hline
\end{tabularx}
\tablefoot{Values in square brackets refer to the modified padding settings adopted for the enlarged $512\times512$ pixel images. }
\end{table*}

\renewcommand{\arraystretch}{1.2}
\begin{table*}[h]
\centering
\caption{Operations in the Inception-ResNet Stem (\texttt{StemRes})}
\label{tab:stem_res}
\begin{tabularx}{\textwidth}{l l c c c c c}
\hline\hline
\textbf{Layer Name} & \textbf{Operation} & \textbf{In Channels} & \textbf{Out Channels} & \textbf{Kernel} & \textbf{Stride} & \textbf{Padding} \\
\hline
\texttt{features[0]} (conv1)    & Conv2d      & \texttt{in\_channels} & 32  & 3$\times$3 & 2 & 0 \\
\texttt{features[1]} (conv2)    & Conv2d      & 32  & 32  & 3$\times$3 & 1 & 0 \\
\texttt{features[2]} (conv3)    & Conv2d      & 32  & 64  & 3$\times$3 & 1 & 1 \\
\texttt{features[3]} (maxpool1) & MaxPool2d   & 64  & 64  & 3$\times$3 & 2 & 0 \\
\texttt{features[4]} (conv4)    & Conv2d      & 64  & 80  & 1$\times$1 & 1 & 0 \\
\texttt{features[5]} (conv5)    & Conv2d      & 80  & 192 & 3$\times$3 & 1 & 0 \\
\texttt{features[6]} (maxpool2) & MaxPool2d   & 192 & 192 & 3$\times$3 & 2 & 0 \\
\texttt{branch\_0}              & Conv2d      & 192 & 96  & 1$\times$1 & 1 & 0 \\
\texttt{branch\_1\_conv1}       & Conv2d      & 192 & 48  & 1$\times$1 & 1 & 0 \\
\texttt{branch\_1\_conv2}       & Conv2d      & 48  & 64  & 5$\times$5 & 1 & 2 \\
\texttt{branch\_2\_conv1}       & Conv2d      & 192 & 64  & 1$\times$1 & 1 & 0 \\
\texttt{branch\_2\_conv2}       & Conv2d      & 64  & 96  & 3$\times$3 & 1 & 1 \\
\texttt{branch\_2\_conv3}       & Conv2d      & 96  & 96  & 3$\times$3 & 1 & 1 \\
\texttt{branch\_3\_avgpool}     & AvgPool2d   & 192 & 192 & 3$\times$3 & 1 & 1 \\
\texttt{branch\_3\_conv}        & Conv2d      & 192 & 64  & 1$\times$1 & 1 & 0 \\
\texttt{concat}                 & Concat      & 96+64+96+64 & 320 & — & — & — \\
\hline\hline
\end{tabularx}
\end{table*}

To investigate the trade-off between model complexity and predictive performance, for both Inception-v4 and Inception-ResNet-v2, we developed several variants that differ primarily in the number of Inception blocks in the network's body. As summarized in Table~\ref{tab:incres}, the final models differ in the number of blocks of each type (A, B, and C), with the Inception-ResNet-v2 model featuring a deeper configuration.

\begin{table}[h!]
  \centering
  \caption{Block structures of the Inception-v4 and Inception-ResNet-v2 models implemented in this work.}
  \label{tab:incres}
  \begin{tabular}{|c|c|}
    \hline
    \rule{0pt}{10pt}\textbf{Inception-v4} & \rule{0pt}{10pt}\textbf{IncResNet-v2} \\
    \hline
    \rule{0pt}{10pt}Input ($299\times299$ image) & \rule{0pt}{10pt}Input ($299\times299$ image) \\
    \hline
    \rule{0pt}{10pt}STEM & \rule{0pt}{10pt}STEM \\
    \hline
    
    \rule{0pt}{10pt}$3\times$ Inception A & \rule{0pt}{10pt}$5\times$ IncRes A \\
    
    \hline
    \multicolumn{2}{|c|}{\rule{0pt}{10pt}Reduction-B} \\
    \hline
    
    \rule{0pt}{10pt}$5\times$ Inception B & \rule{0pt}{10pt}$10\times$ IncRes B \\
    
    \hline
    \multicolumn{2}{|c|}{\rule{0pt}{10pt}Reduction-A} \\
    \hline
    
    \rule{0pt}{10pt}$3\times$ Inception C & \rule{0pt}{10pt}$5\times$ IncRes C \\
   
    \hline
    \rule{0pt}{10pt}Average Pooling & \rule{0pt}{10pt}Average Pooling \\
    \hline
    \multicolumn{2}{|c|}{\rule{0pt}{10pt}FC-1536 $\rightarrow$ FC-768 $\rightarrow$ FC-384 $\rightarrow$ FC-192 $\rightarrow$ FC-96} \\
    \hline
    \multicolumn{2}{|c|}{\rule{0pt}{10pt}FC-5} \\
    \hline
  \end{tabular}
\end{table}

A critical adaptation for our regression task was modifying the final classifier head. We replaced the original global average pooling and softmax layer with a sequence of FC layers that progressively reduce the feature vector dimensionality down to the final 5-parameter output. This final FC block also includes dropout modules (with a keep probability of $p=0.3$) after the first three FC layers to regularize the network. A Leaky-ReLU activation function is used in the final layer.

\end{appendix}

\label{LastPage}
\end{document}